\documentclass[3p,11pt]{elsarticle} 
\usepackage{algorithm}
\usepackage{algpseudocode}
\usepackage{graphicx}
\usepackage{comment}
\usepackage{amsmath}
\usepackage{subfigure}

\usepackage{booktabs} 
\usepackage{color}
\usepackage[english]{babel}
\usepackage{picinpar}
\usepackage{wrapfig}
\usepackage{graphicx}
\usepackage{moresize}
\usepackage{amssymb}
\usepackage{url}
\usepackage{subfigure}
\usepackage{amsmath}

\usepackage{algorithm}

\DeclareMathOperator*{\argmin}{arg\,min}
\newcommand{\mt}[1]{\textcolor{black}{#1}}

\begin{document}
\begin{frontmatter}

\title{DASC: Towards A Road Damage-Aware Social-Media-Driven Car Sensing Framework for Disaster Response Applications}
\author{Md Tahmid Rashid}
\author{Daniel (Yue) Zhang}
\author{Dong Wang}
\address{Department of Computer Science and Engineering\\
University of Notre Dame\\
Notre Dame, IN 46556}
\begin{abstract}
While vehicular sensor networks (VSNs) have earned the stature of a mobile sensing paradigm utilizing sensors built into cars, they have limited sensing scopes since car drivers only opportunistically discover new events. Conversely, social sensing is emerging as a new sensing paradigm where measurements about the physical world are collected from humans. In contrast to VSNs, social sensing is more pervasive, but one of its key limitations lies in its inconsistent reliability stemming from the data contributed by unreliable human sensors. In this paper, we present DASC, a road \textbf{D}amage-\textbf{A}ware \textbf{S}ocial-media-driven \textbf{C}ar sensing framework that exploits the collective power of social sensing and VSNs for reliable disaster response applications. However, integrating VSNs with social sensing introduces a new set of challenges: i) How to leverage noisy and unreliable social signals to route the vehicles to accurate regions of interest? ii) How to tackle the inconsistent availability (e.g., churns) caused by car drivers being rational actors? iii) How to efficiently guide the cars to the event locations with little prior knowledge of the road damage caused by the disaster, while also handling the dynamics of the physical world and social media? The DASC framework addresses the above challenges by establishing a novel hybrid social-car sensing system that employs techniques from game theory, feedback control, and Markov Decision Process (MDP). In particular, DASC distills signals emitted from social media and discovers the road damages to effectively drive cars to target areas for verifying emergency events. We implement and evaluate DASC in a reputed vehicle simulator that can emulate real-world disaster response scenarios. The results of a real-world application demonstrate the superiority of DASC over current VSNs-based solutions in detection accuracy and efficiency.
\end{abstract}
 
 \begin{keyword}
vehicle sensor networks, social sensing, bottom-up game theory, incentive control, Markov Decision Process.
\end{keyword}

\end{frontmatter}
	\section{Introduction} \label{sec:intro}
Vehicular sensor networks (VSNs) have evolved into a robust networked sensing paradigm for obtaining situational awareness in disaster response applications~\cite{zhang2008efficient}. VSNs incorporate cars equipped with arrays of on-board sensors (e.g. dashboard cameras) to opportunistically identify event occurrences like gas unavailability at nearby gas stations or accidents on the roads \cite{park2016motives}. Social sensing, on the other hand, is a permeating sensing paradigm for collecting real-time measurements about the physical world from observations reported by social media users \cite{wang2019age}. Examples of social sensing applications include monitoring air quality in smart cities~\cite{zhang2018real}, studying human mobility in urban areas~\cite{noulas2012tale}, and obtaining situation awareness in the aftermath of disasters~\cite{zhang2017constraint} using online social media (e.g., Twitter, Instagram). 

While VSNs render greater reliability in the discovery of the ground truth of events using physical sensors, one limitation is that the information collected by the vehicles is restricted to only those regions traversed by car drivers. Such a limitation vastly limits the scope of sensing for VSNs and their adaptability in unraveling new events. Moreover, during a disaster situation, roads could become inaccessible due to damages caused by the disaster, rendering VSNs ineffective in certain scenarios. In contrast to VSNs, the scale of social sensing is broader and any individual possessing a smart device with Internet connectivity can potentially report an event on social media. However, an inherent limitation of social sensing is the inconsistent reliability of the sensing data that are often contributed by unreliable human sensors~\cite{wang2015social}. In this paper, we exploit the complementary nature and collective strengths of VSNs and social sensing to develop DASC, a Damage-Aware Social-media-driven Car sensing framework.

\begin{wrapfigure}{l}{0.61\textwidth}
    \centering
    \includegraphics[width=10cm]{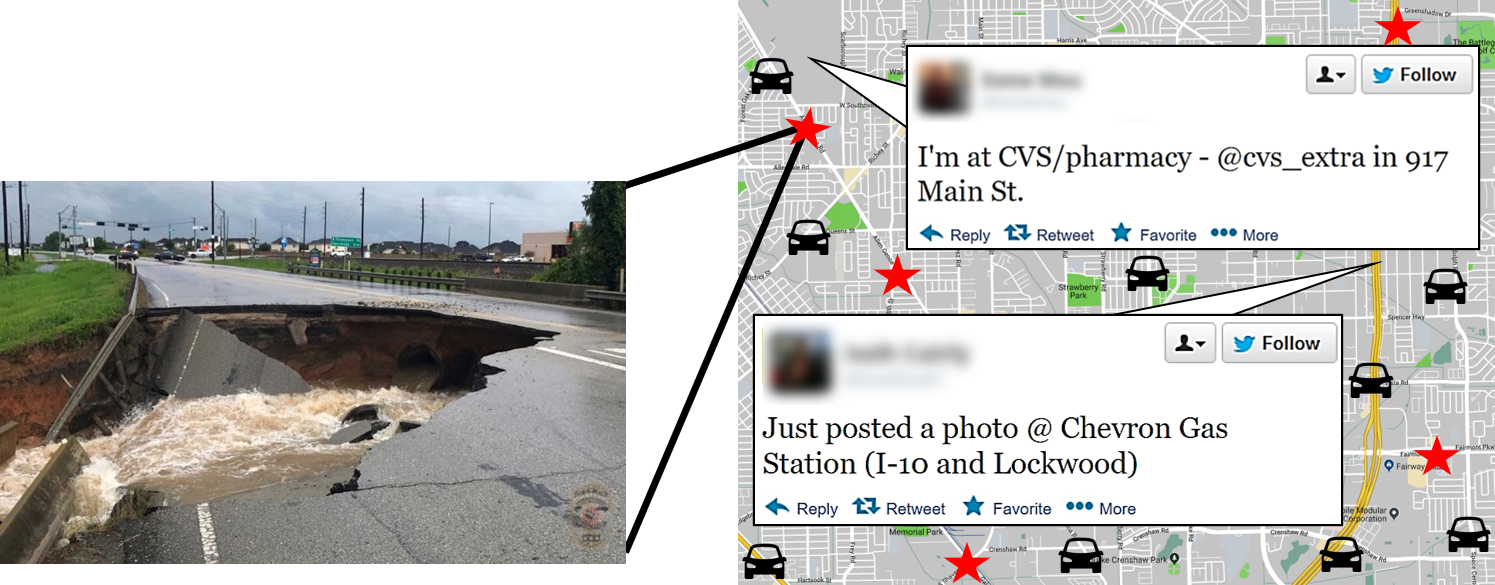}
    \caption{Tweets Posted During Hurricane Harvey}
    \vspace{-0.1in}
    \label{fig:Tweets}
\end{wrapfigure}

Consider Hurricane Harvey that occurred in Southern Texas in August 2017 as an example. In the aftermath of the hurricane, access to critical resources (e.g., gas stations, pharmacies, food) became crucial for the victims affected by the disaster~\cite{lichtveld2018disasters,zhang2019crowdlearn}. Figure \ref{fig:Tweets} shows tweets posted during Hurricane Harvey. If these tweets could be utilized to direct car drivers to desired locations, the disaster recovery process could be facilitated by obtaining facts regarding the reported events using vehicular sensors. However, a few key technical challenges need to be addressed for developing a reliable social-media-driven car sensing system.

The first challenge is leveraging the sparse and unreliable social media data to guide cars to desired locations. A key challenging task in social sensing applications is the accurate identification of reliable sources and truthful claims from the sparse and uncertain social sensing data, otherwise known as \textit{truth discovery}~\cite{wang2019social}. To discover truthful information from unvetted social media users, existing truth discovery solutions primarily rely on the posts presented on social media. These solutions may yield unreliable truth discovery results, making it difficult to decide where to dispatch the cars \cite{zhang2016robust}. In addition to that, existing truth discovery algorithms that output probabilistic distributions for the classification cannot definitively confirm the truthfulness of an event. Therefore, it is intrinsically difficult to extract reliable social signals to guide cars to accurate locations of interests. We deem this challenge as the \textit{cyber challenge} of the problem.

The second challenge is the inconsistent availability, otherwise known as \textit{churn}, caused by the rational car drivers who may drop sensing tasks midway during explorations. We assume that at times, car drivers may focus on personal goals instead of the given assignment and abandon the sensing tasks abruptly in the middle of a trip. This inconsistent availability issue of \textit{churn} has been explored in the field of distributed systems \cite{rhea2004handling, vance2019towards,zhang2019heteroedge}. Existing literature has proposed methods to reduce churns by increasing the total number of devices in the system \cite{godfrey2006minimizing, haeberlen2006fallacies} or by reallocating tasks to more reliable devices~\cite{zhang2010cloud}. While those solutions may work for networked devices, they are application-specific and are hard to be applied to cars. For example, it may not be possible to rigorously control the number of cars in the system since the cars are privately owned by individuals. Moreover, while allocating tasks to more reputable drivers (i.e., drivers who are less likely to drop a task) may improve the performance of the system, it could be impractical to switch the tasks to a car that is in the middle of an ongoing exploration on the road. Hence, it remains an open challenge to decisively allocate sensing tasks to cars (drivers) in our DASC framework. We deem this challenge as the \textit{human challenge} of the problem.

\begin{wrapfigure}{l}{0.35\textwidth}
    \centering
   \vspace{-0.05in} 
    \includegraphics[width=6cm]{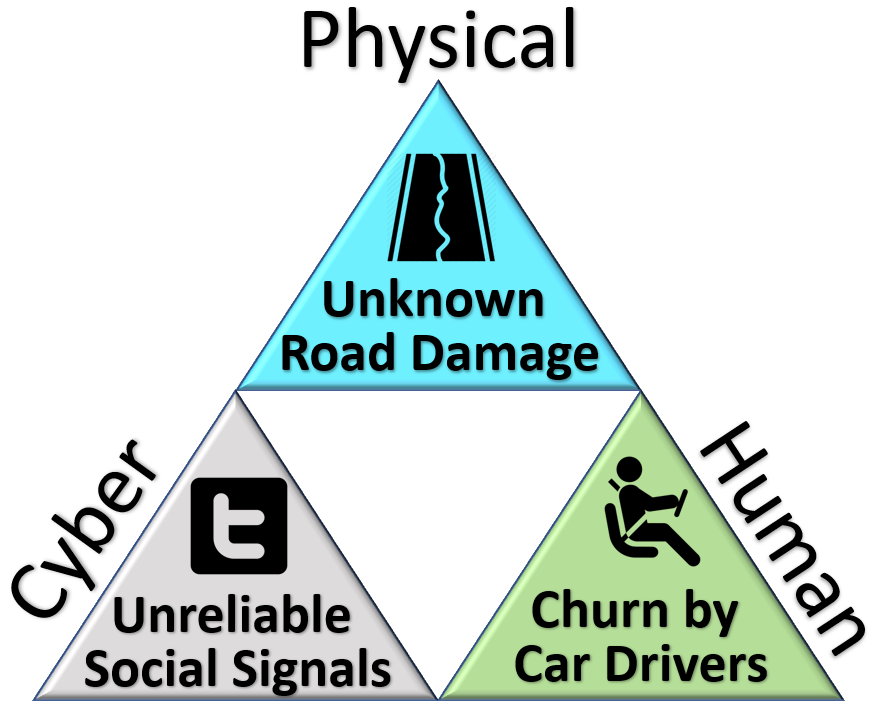}
    \vspace{-0.25in}
    \caption{The three challenges in social-media-driven car sensing system}
    \vspace{-0.05in}
    \label{fig:threedim}
\end{wrapfigure}

The third challenge is allocating the cars to the event locations with little or no prior knowledge of road damages caused by the disaster. After a major natural disaster, it is likely that a certain proportion of roadways are unreachable \cite{bono2011network}, as exemplified in Figure \ref{fig:Tweets} (marked by red stars). Road damages inflicted by the disaster would greatly limit the maneuverability of the cars. Furthermore, the extent of road damages is unpredictable and cannot be known beforehand. A few routing strategies have been proposed to model the road damages and route cars safely \cite{baker2003genetic, prins2004simple}. However, existing solutions assume that the system has global knowledge of all the road damages, making feasible routing decisions. In contrast, with damaged infrastructures after a disaster, the information about the road conditions cannot be readily determined and disseminated. For example, at any point in time, new road damages may appear and also existing road may get repaired, introducing a level of dynamism into the system. Current solutions in the discipline of routing algorithms consider the issue of unexplored and incomplete information. However, in the context of social-media-driven car sensing systems, the dynamics of social media combined with that of the physical world make the problem more challenging. We deem this challenge as the \textit{physical challenge} of the problem. Figure \ref{fig:threedim} illustrates the three challenges in developing a social-media-driven car sensing system.

In this paper, we develop DASC, a damage-aware social media-driven car sensing system to address the above challenges. To address the first challenge, we develop a bottom-up game-theoretic task allocation model to judiciously dispatch the cars to reported locations and verify the event information extracted from unreliable social media data. To address the second challenge, we design a top-down incentive control mechanism to dynamically adjust the incentives for exploration of the event locations based on the aggregated reputations of the cars (i.e., the historical behavior of the cars in attempting to successfully complete the tasks). To address the third challenge, we develop a Markov Decision Process (MDP)-based damage discovery scheme to locate the roads affected by damage and leverage the obtained knowledge to make optimal route planning decisions. To the best of our knowledge, DASC is the first solution  that melds the realms of social sensing and VSNs for a robust car sensing (SCS) application with explicit consideration of the road damages in the aftermath of a disaster. We implemented the DASC framework and evaluated it with CARLA, a reputed car emulation system that is capable of closely replicating road networks and vehicles in real-world scenarios. We compared our framework against representative vehicular based sensing systems on a real-world dataset collected from Twitter during a natural disaster: Hurricane Harvey in August 2017. The results show that DASC significantly outperforms the compared baselines in both detection effectiveness and deadline hit rate during the aftermath of a disaster.

A preliminary version of this work has been published in~\cite{rashid2019socialcar}. We refer to the scheme developed in the prior work as the SocialCar scheme. This paper is a significant extension of the previous work in the following aspects. First, we define a new problem of guiding cars to the event locations using social media signals while explicitly considering the damaged roads left by a disaster. In contrast to the SocialCar scheme, the new problem is much more challenging because immediately after the disaster, the knowledge of the road damage is not available, making it much more difficult to decide the route for sending the cars to the event locations. \mt{Second, we allocate a portion of participating cars for ``scouting" the routes to explore the damaged roads.} Third, we introduce a Markov Decision Process (MDP)-based technique for utilizing the knowledge of the road damage obtained in the prior step for feasible route planning. Fourth, we employ an exploration-exploitation strategy to learn the optimal routing decision over time. Fifth, we carry out a new set of experiments to explicitly evaluate the performance of all schemes in terms of detection effectiveness and deadline hit rates in the new problem setting. Sixth, we include additional baselines to further exhibit the performance gains achieved by DASC. Finally, we extend the related work by adding a new discussion on the damage-aware routing schemes and highlight the difference between DASC and those schemes (Section~\ref{sec:related}).
	\section{Related Work}\label{sec:related}

\subsection{Vehicular Network Based Sensing}
The emergence of modern cars equipped with advanced sensors has opened new domains in vehicular networked sensing \cite{park2016motives}. For example, Nekovee presented a comprehensive study of several roadside monitoring systems that integrate the data of nearby vehicular sensors \cite{nekovee2005sensor}. Lee \textit{et al.} proposed Mobeyes, an urban surveillance system that manages a dedicated number of sensor-equipped cars to opportunistically explore events~\cite{lee2006mobeyes}. In combination with built-in sensors, crowdsourced VSNs using smartphones have also gained popularity. Commercial solutions like Waze \cite{galeso2016waze} and GasBuddy \cite{dong2008automatic} provide users with valuable information about the availability of critical resources as reported by car drivers. However, the above approaches primarily depend on the opportunistic nature of the car drivers since cars only ``sense" incidents when they come across them. In contrast, this paper presents the DASC framework that leverages the social media signals to guide cars to desirable locations for a greater sensing scope.

\subsection{Social Sensing}
Social sensing is transcending as a recent sensing paradigm that uses humans as sensors to report about observations in the physical world \cite{IPSN:12,wang2014using}. Examples of social sensing applications include  identifying traffic risks~\cite{zhang2020graphcast},  tracking social unrest~\cite{al2014crowd} and disasters~\cite{marshall2016mood,wang2013recursive}, sensing points of interest in large cities~\cite{zhang2019sparse,zhang2017large}, and detecting car plates of suspects \cite{zhang2019edgebatch,zhang2019heteroedge}. Several challenges have been studied in social sensing. Examples include data reliability \cite{wang2013credibility}, human uncertainty \cite{wang2012scalability}, data sparsity \cite{zhang2019deeprisk}, privacy preservation \cite{vance2018privacy}, and real-time requirements \cite{zhang2019integrated}.  A comprehensive survey of social sensing schemes is provided in~\cite{wang2015social}.
Xu \textit{et al.} developed a framework for semantic and spatial analysis of urban emergency events using social sensing \cite{xu2016participatory}. Chen \textit{et al.} proposed a road traffic congestion monitoring system using social media data \cite{chen2014road}. Imran \textit{et al.} developed a machine learning-based disaster identification system capable of classifying and analyzing information from crisis-related tweets in real-time \cite{imran2014aidr}. A key limitation of current social sensing systems is that they only rely on the social media data which could be unreliable~\cite{wang2014provenance,wang2014surrogate,zhang2019social}. More recently,  there is an inception of social media driven UAV-based sensing approaches that address the data reliability issue of social sensing by using physical drones~\cite{rashid2019collabdrone,rashid2019sead}. However, these solutions require dedicated drones for the sensing purpose that are typically known to be expensive and limited in numbers. In contrast, the SocialCar framework integrates social media with existing vehicular-based sensing systems to provide data reliability assurance in scalable social sensing.

\subsection{Mitigation of High Churn}
Recent literature has presented methods to diminish the issue of high churn in participatory sensing and distributed systems. For example, Gao \textit{et al.} \cite{gao2015survey} presented a study of different incentive mechanisms for participatory sensing to lower the possibility of churn. Godfrey \textit{et al.} \cite{godfrey2006minimizing} proposed a technique to reduce churn by intelligently selecting only the most reliable devices from all participating devices in a system. Haeberlen \textit{et al.} proposed a solution to reduce churn by increasing the total number of existing devices \cite{haeberlen2006fallacies}. One major drawback of these approaches is that they are all either bottom-up or top-down and do not holistically consider the objectives of the individual devices and the server. In contrast, our DASC framework uses both a bottom-up task preference and a top-down dynamic incentive control to collectively consider the objectives on both ends (i.e. the cars and the SVS application) to better control the churn issue.

\subsection{Road Damage-aware vehicle routing}
Road damage-aware vehicle routing is a well-studied topic in vehicular networks. For example, Hsueh \textit{et al.}~\cite{hsueh2008dynamic} presented a comprehensive study of road damage-aware dynamic vehicle routing for relief logistics during natural disasters. Korkmaz \textit{et al.}~\cite{korkmaz2016path} proposed a road damage-aware satellite-imagery based path planning framework for rescue vehicles during disasters. Mahmoudabadi \textit{et al.}~\cite{mahmoudabadi2014solving} developed a damage severity-aware route planning for transporting hazardous substances during emergencies. Kuntze \textit{et al.}~\cite{kuntze2012seneka} explored the possibility of collision-free path planning of unmanned ground vehicles (UGVs) in road-damage prone areas. While the above approaches intend to solve the critical challenge of guiding vehicles through road damage, our problem of building a damage-aware social-media-driven car sensing system is even more challenging due to both the dynamics of the social media and the physical world. In this paper, we develop the DASC framework to address this challenge by designing an MDP-based damage discovery technique to locate damaged roads and use the knowledge for optimal routing.
	\newtheorem{myDef}{DEFINITION}
\section{Problem Formulation} \label{sec:problem}
In  this section, we present the fundamental definitions and assumptions of our model and define the objective of our problem. In a damage-aware social-media-driven car sensing (SCS) application, we inspect a physical region of interest (ROI) for a specific duration of sensing. The sensing timeline is discretized into $T$ periodic intervals, namely \textit{response cycles}. In particular, $t \in [1,T]$ indicates the $t^{th}$ response cycle.

\mt{\begin{myDef}
    \emph{\textbf{Sensing Cells:} We divide the sensing region reachable by cars into $H$ disjoint \textit{sensing cells}. Each cell represents a real-world location, connected by roads and accessible to cars~\footnote{Please note that we only focus on the events in the sensing cells defined above and ignore the events that happened in locations which are not accessible to cars given the scope of this work.} At any given time, a sensing cell can be occupied by multiple cars, increasing the chance of a task being completed. In particular, we define $SC_{t,h}$ to be the $h^{th}$ sensing cell at the $t^{th}$ response cycle.}
\end{myDef}}

\begin{myDef}
    \emph{\textbf{Road Damage $D_{t,h}$ for sensing cell $SC_{t,h}$:}
We assume that cars cannot traverse a sensing cell if it contains road damage. We identify a sensing cell's damage state by a binary variable $D_{t,h}$, where a value of 0 indicates no damage and a value of 1 indicates damage.
}
\end{myDef}

We consider that a set of social media users reports a collection of independent \textit{events} at different sensing cells as defined below:
\begin{myDef}
    \emph{\textbf{Event $E_{t,n}$:} An \textit{event} is assumed to represent a physical variable of interest within a sensing cell in the SCS application. Examples of reported events include gas availability at a gas station or a person trapped under a vehicle. We let a binary variable $E_{t,n}$ denote the $n^{th}$ event in the $t^{th}$ response cycle with a total of $N_t$ events. For each reported event $E_{t,n}$, it either ``exists" (i.e., $E_{t,n} = 1$) or ``does not exist" (i.e., $E_{t,n} = 0$).
}
\end{myDef}

We use $\widehat{E_{t,n}}$ to denote the truth of event ${E_{t,n}}$ estimated by our DASC system. An essential attribute of each event is the \textit{sensing deadline} as defined below:

\mt{\begin{myDef}
	\emph{\textbf{Sensing Deadline $\delta_{t,n}$ for event $E_{t,n}$:} Each event is assigned a deadline representing the urgency by analyzing the content of the social media report~\cite{singh2018analyzing}. We assume that the deadlines of events are shorter than the duration of the sensing cycle~\cite{zhang2019edgebatch}. For an event with a longer deadline than the sensing cycle, we split the event into multiple events, each resulting event having a deadline shorter than the sensing cycle. The split events also inherit the \textit{priority} from the original event to avoid the potential problem of ``priority inversion"~\cite{rashid2019sead}.}
\end{myDef}}

We define the data from social media (e.g., tweets from Twitter) as follows:
\begin{myDef}
	\emph{\textbf{Social Media Data $\mathcal{S}$:} A set of social media posts that reports events in the physical world in the aftermath of a disaster. An example of such a report is shown in Figure~\ref{fig:Tweets}.}
\end{myDef}

We consider a set of \textit{tasks} that are broadcast for the car drivers to explore the reported events. We formally define a task as:

\begin{myDef}
	\emph{\textbf{Task $V_{t,n}$:} A task for a car at each response cycle refers to the location of an event (i.e., a sensing cell) where the car should be dispatched to investigate the event. 
	}
\end{myDef}

\begin{wrapfigure}{r}{0.61\textwidth}
    \centering
    \includegraphics[width=9.5cm]{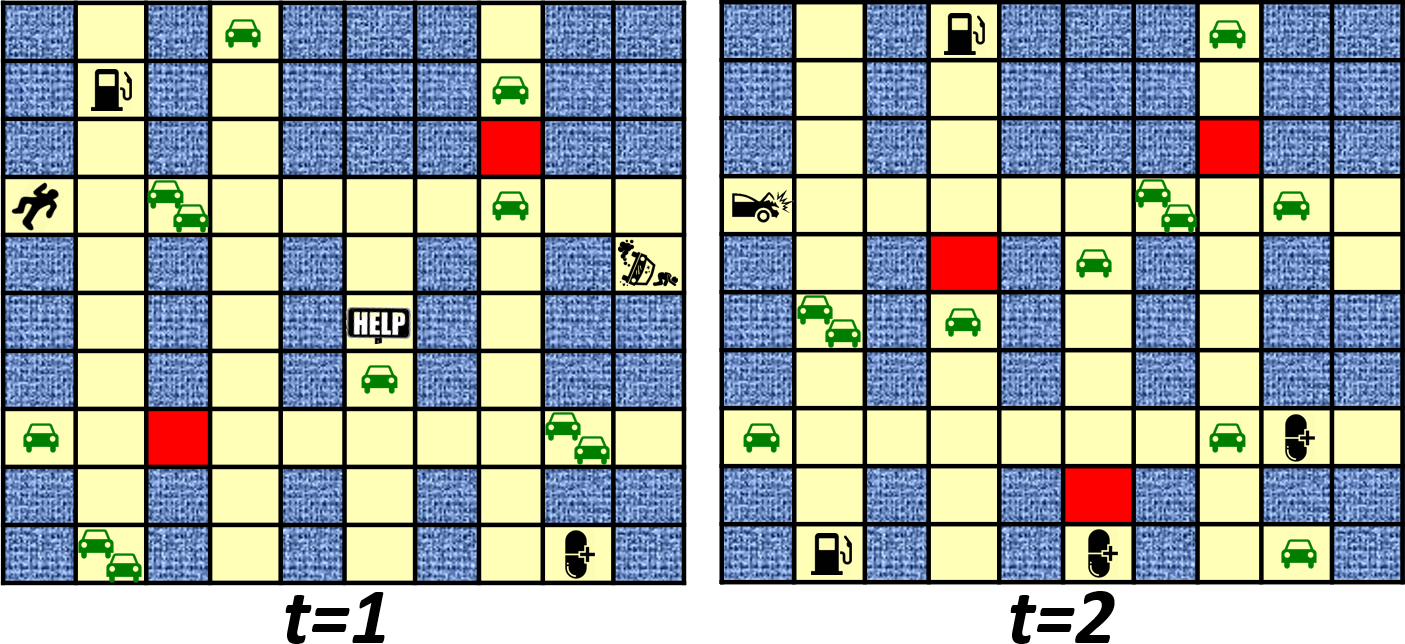}
    \vspace{-0.25in}
    \caption{Snapshot of the sensing grid across response cycles. The yellow boxes signify \textit{sensing cells}, the blue boxes represent unreachable regions, the red boxes signify the cells with road damage, the black icons indicate the events, and the green car icons represent the cars.}
    \vspace{-0.10in}
    \label{fig:grid}
\end{wrapfigure}

Figure~\ref{fig:grid} presents an illustrative example of the concepts defined above.


We make two important assumptions about the unique compliance and churn issues in the DASC system.

\emph{\textbf{Voluntary Compliance}}: We assume that a car driver may or may not be willing to accept any task offered.

\emph{\textbf{Dynamic Churn}}:  We also assume that even if car drivers are willing to pick up events for investigation, they may randomly abort the tasks (e.g., the driver decides to go home after taking a task), causing the churn in the system.

Using the above definitions, we, therefore, define the goal of our DASC framework. Given the social media data input $\mathcal{S}$, a set of cars $\mathcal{C}$, the corresponding deadlines for the events $\delta_{t,n}$, the road damage $\mathcal{D}$, as well as sensing cells $SC_{t,h}$, the objective of the DASC framework is to dispatch the cars to a set of sensing cells to maximally recover the true states of the events reported by social media users while considering the road damage caused by the disaster. 
We formally solve a constrained optimization problem as follows:
\begin{equation}\label{eq:deadlinecover}
 \begin{split}
 \argmin_{\widehat{E_{t,n}}} &\sum_{n=1}^{M_t} (abs(\widehat{E_{t,n}}-E_{t,n}) | \mathcal{D}, \mathcal{S}, \mathcal{C}, \delta_{t,n}, SC_{t,h}), \\&~\forall 1 \le t \le T,~\forall 1 \le h \le H\\
\end{split}
\end{equation}\noindent

The definitions of the notations are summarized in Table~\ref{tab:notations}.
    \begin{table}[!h]
    \caption{Summary of Notations}
    \centering
    \begin{tabular}{|l|l|}
        \hline
        $t$  & The $t^{th}$ response cycle, $t\in \{1,2,...,T\}$\\ \hline
        $E_{t,n}$  & The $n^{th}$ event in response cycle $t$\\ \hline
        $\widehat{E_{t,n}}$ & The estimated truth of event $E_{t,n}$\\ \hline
        $\delta_{t,n}$ & The sensing deadline for event $E_{t,n}$\\ \hline
        $SC_{t,h}$ & The $h^{th}$ sensing cell in response cycle $t$\\ \hline
        $D_{t,h}$ & The road damage for sensing cell $SC_{t,h}$\\ \hline
        $V_{t,n}$ & The $n^{th}$ task in response cycle $t$\\ \hline
        $\mathcal{S}$ & The social media data\\ \hline
        $C_b$ & The $c^{th}$ car, $c\in \{1,2,...,G\}$\\ \hline
    \end{tabular}
    \label{tab:notations}
    \end{table}

	\section{The DASC Framework}
\begin{wrapfigure}{l}{0.46\textwidth}
    \centering
    \vspace{-0.05in}
    \includegraphics[width=8cm]{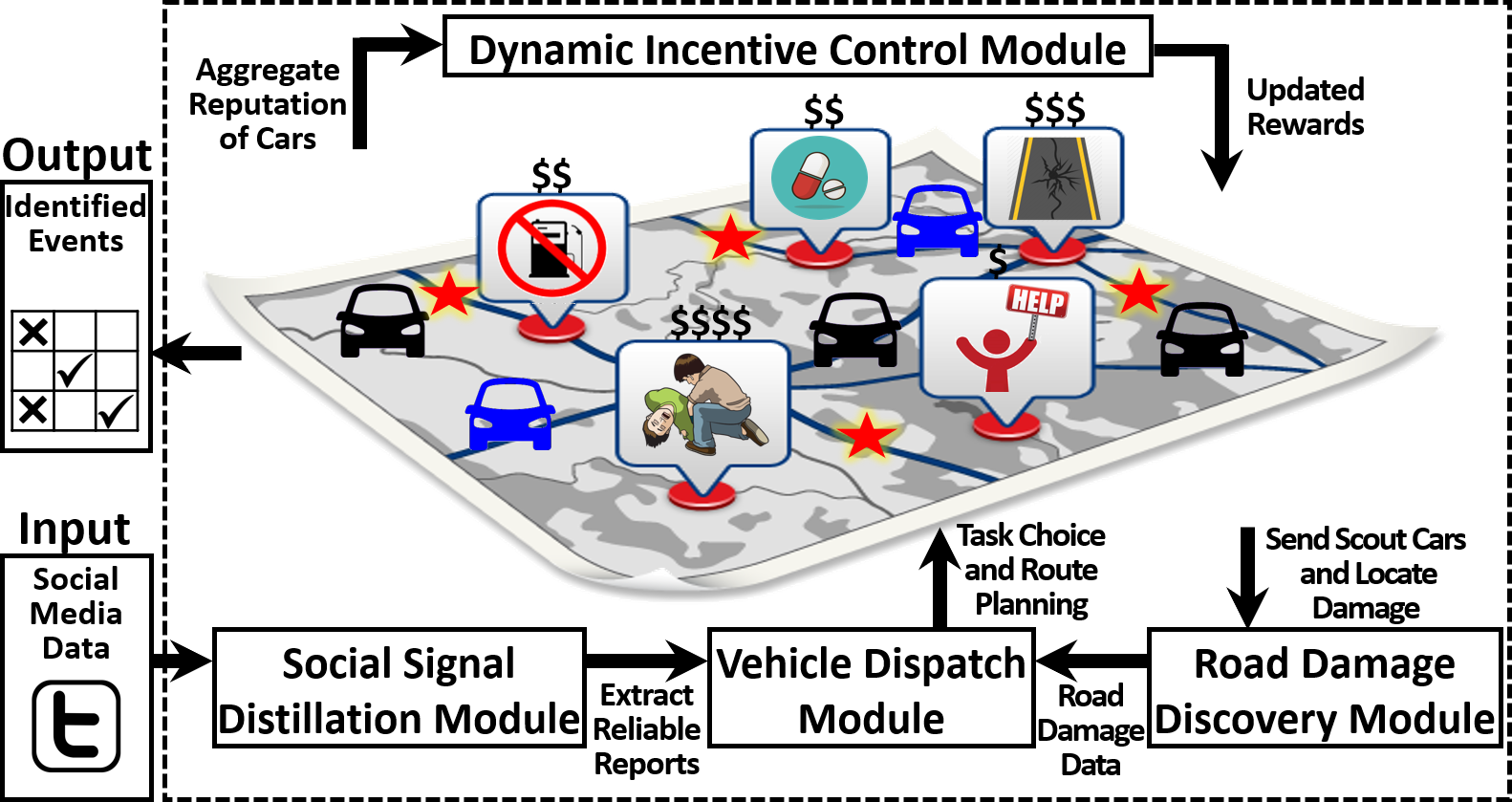}
    \vspace{-0.35in}
    \caption{Overview of the DASC Framework. The blue car icons represent the cars sent out for locating the road damage, the black car icons represent the cars dispatched to event locations, and the red stars indicate the discovered road damages along the routes.}
    \vspace{-0.1in}
    \label{fig:Framework}
\end{wrapfigure}

In this section, we present the DASC framework that integrates the social media and the vehicular sensing system for a reliable road damage-aware SCS application. An overview of DASC is shown in Figure~\ref{fig:Framework}. The DASC consists of four major components: i) a Social Signal Distillation (SSD) module; ii) a Road Damage Discovery (RDD) module; iii) a Vehicle Dispatch (VD) module ; and iv) a Dynamic Incentive Control (DIC) module.

The SSD module collects and extracts reliable event reports from unreliable social media users. Concurrently, the RDD module assigns a portion of the participating cars as scout cars to explore the routes to the event locations for road damage. The VD module allows the car drivers to pick their preferred tasks based on their individual payoffs and leverages the knowledge from road damage data to guide the cars to the event locations. The DIC module assigns and adjusts the incentives for the tasks to maximize the chance of the cars completing all the tasks (i.e. locating road damage and exploring all the reported events). To further maximize the chance of the tasks being completed, once each task is selected by at least one car driver, the VD module allows the tasks to be selected by multiple car drivers. A detailed discussion of each module is presented in the following subsections.

    \subsection{Social Signal Distillation (SSD) Module}
    The SSD module is designed to collect, pre-process, and analyze noisy social media posts to estimate the possibility of critical events in the physical world. The SSD module uses a real-time data crawler engine to obtain social media data (e.g. Tweets) with geo-location tags indicating disaster-related events. The collected data is filtered by running keyword searches on it (e.g., gas, fuel, oil, medicine, healthcare, and pharmacy), and afterward clustered and labeled using a microblog data clustering tool~\cite{zhang2016robust}. A key issue with the above generated social media data lies in the trustworthiness of the reported events since these events  are often reported by unvetted grass-root users, of whom the credibility is unknown a priori. Without carefully excluding the misinformation and rumors provided by unreliable users, the performance of the DASC system can be significantly degraded. Another frequent issue observed in social media data is “data sparsity”, whereby a majority of the users contribute only a small number of event reports, providing insufficient evidence to accomplish the truth estimation task. In light of such challenges, the SSD module incorporates a truth discovery (TD) solution to estimate the truthfulness of the reported events along with obtaining the estimation confidence/uncertainty~\cite{zhang2020pqa}.
	
	While there is an abundant number of TD solutions, we select a particular approach called the Robust Truth Discovery (RTD)~\cite{zhang2016robust} algorithm for our SSD module to filter useful signals from social sensing data. Our main motivation for selecting this algorithm lies in its design philosophy to be robust against misinformation spread and data sparsity in social media applications. In particular, the RTD scheme handles widespread misinformation by explicitly quantifying different degrees of attitude that a source may express on a claim and incorporating the historical contributions of a source using a principled approach. The fine-grained source attitude facilitates an effective detection of misinformation, which is based on the observation that the misinformation is more likely to attract opposite opinions and intensive debates. Moreover, the RTD scheme addresses the data sparsity issue by computing the claim truthfulness based on a function of the source attitude, the source’s historical contributions, and the source reliability. The estimation is more robust since it does not solely rely on the source reliability estimation, which is challenging to estimate accurately in a sparse dataset~\cite{zhang2019sparse}. The RTD scheme measures the historical claims of each source by computing a metric called the \textit{contribution score} that determines a source's contribution to an event report based on several factors~\cite{zhang2016robust}. The algorithm also utilizes a metric called the \textit{source attitude score} to fully capture the reporting behavior of sources. We define the output of the RTD algorithm as \textit{event veracity} and \textit{estimation confidence} which are defined below:
	
	\begin{myDef}
        \emph{\textbf{Event Veracity $\Lambda_{t,n}$ for event $E_{t,n}$:} A score in the range (0,1] that indicates the chance of an event being true. Intuitively, the greater the value of $\Lambda_{t,n}$, the more likely event $E_{t,n}$ is true (i.e., $E_{t,n}$ exists). To obtain $\Lambda_{t,n}$ for event $E_{t,n}$, the RTD algorithm iteratively sums up all the contribution scores from the set of social media users who contribute to $E_{t,n}$.}
    \end{myDef}

    \begin{myDef}\label{confidence}
        \emph{\textbf{Estimation Confidence Score $EC_{t,n}$ for event $E_{t,n}$:}  A score in the range (0,1] that signifies the estimation confidence for an event. Intuitively, the greater the value of $EC_{t,n}$, the more confident the RTD algorithm is its estimation. Formally, it is defined as the absolute difference between the event veracity score and the midpoint of event veracity score's range (i.e., the neutral point for determining the truthfulness).}
    \end{myDef}

    Leveraging social sensing and truth discovery, the SSD module of the DASC framework not only provides the signals of the critical events for the vehicles to verify but also helps to quantify the priorities of these events based on their confidence. Both the event truthfulness and the estimation confidence are critical inputs to the DASC framework that guide the dispatching strategies of the vehicles in the VD module. The decision to dispatch the cars rely on the values of $\Lambda_{t,n}$ and $EC_{t,n}$. For a given response cycle, if the value of $EC_{t,n}$ is above an adjustable threshold, DASC trusts the RTD algorithm's decision without dispatching the vehicles and concludes upon event $E_{t,n}$'s truthfulness based on the value of $\Lambda_{t,n}$. For cases otherwise, where the value of $EC_{t,n}$ is below the threshold deeming the veracity doubtful, DASC incorporates the value of $EC_{t,n}$ into the VD module (the process of which is detailed in Section 4.3) and allocate tasks for the car drivers to explore the event. Once the cars travel to the event destination and collect the actual truth with greater reliability using the onboard sensors, DASC finally determines event $E_{t,n}$'s truthfulness.

\subsection{Road Damage Discovery (RDD) Module}
The RDD module is designed to incentivize and assign a fraction of cars (from the pool of available cars that are willing to participate) to explore the available routes for road damages. In particular, we assign $Q\%$ of all the available cars, that are willing to participate in the sensing process, as \textit{scout cars}, where $Q$ is adjusted according to the application scenario. We model each road intersection as the \textit{node} of a graph and each road branching out of an intersection as the \textit{edge} of a graph. Each exploration is modeled as a \textit{task} that can be picked up by one or multiple cars and assigned a \textit{reward} $r_{t,n}$ to incentivize the drivers to pick it up. 
The \textit{reward} $r_{t,n}$ is determined by the Dynamic Incentive Control (DIC) module discussed later in this section. The rationale is that if we can traverse the maximum number of roads using the scout cars, we may have a better chance of locating road damages. 

Once a scout car is adjacent to a cell with road damage, the damage information (i.e., $D_{t,h}$) is recorded by the scout car and then it proceeds to explore a different route. The edges of the graph are basically a series of contiguous sensing cells accessible by cars. While the road damage variable $D_{t,h}$ indicates whether a cell has damage or not, we acknowledge that all the road damage information cannot be readily obtained or updated in a given sensing cycle. Therefore, it is a reasonable assumption to determine the possibility of road damage across a sensing cell based on the historical damage condition of the cell. In order to accomplish this, all the sensing cells that make up the edges of the graph are assigned a score called \textit{accessibility index}, which is defined below.
\begin{myDef}
\emph{\textbf{Accessibility Index $X_{t,h}$:} A score in the range of $[0,1]$ to indicate the possibility of road damage across a sensing cell (i.e., how likely damage may occur again in a cell in the future). Intuitively, a lower accessibility index indicates that a route is less likely to be traversable due to the possibility of containing damaged roads. Initially, all the cells are considered to have an initial accessibility index $X_{0,h}$, the value of $X_{0,h}$ is discussed in Section \ref{sec:eval}. Over response cycles, only when a sensing cell is visited by a scout car, the accessibility index is calculated as:}
\end{myDef}

\begin{equation}\label{eq:access}
X_{t,h}=
\begin{cases}
X_{t-1,h}-\kappa_t, & D_{t,h}=1\\
X_{t-1,h}+\kappa_t, &\text{otherwise}\\
\end{cases}, 0\leq X_{t,h}\leq1
\end{equation}
where $\kappa_t$ is an adjustable parameter called accessibility penalty, which is determined by a sliding window correlation~\cite{mokhtari2019sliding} between the total number of detected road damages in $t^{th}$ sensing cycle, $D^{total}_{t}$ and the total number of reported events in $t^{th}$ sensing cycle, $N_t$. The value of $\kappa_t$ is computed by the following equation:
\begin{equation}\label{eq:kappacalc}
\kappa_t=\frac{\sum_{i=t-j+1}^t\{(D^{total}_{t}-\widehat{D^{total}_{t}})\times(N_t-\widehat{N_t})\}}{\sqrt{\sum_{i=t-j+1}^t(D^{total}_{t}-\widehat{D^{total}_{t}})^2\times \sum_{i=t-j+1}^t {(N_t-\widehat{N_t})^2}}}
\end{equation}
\noindent

where $j$ is the sliding window of the number of cells to look back, $\widehat{D^{total}_{t}}$ is the average number of road damages during the sliding window, and $\widehat{N_t}$ is the average number of events during the sliding window. Intuitively, if there is a strong correlation between the number of events and the road damages in the current sensing cycle, the value of $\kappa_t$ will increase, thereby making the accessibility index more sensitive.

In the beginning, the road damage $D_{t,h}=0$ for all the cells, assuming that all the roads are traversable. If a cell is not visited by a car in a sensing cycle, the accessibility index for that cell is retained (i.e. $X_{t,h}=X_{t-1,h}$). If a cell is found to have damage at a sensing cycle (i.e. $D_{t,h}=1$), the accessibility index is decremented by $\kappa$. Conversely, if a cell is detected to have no damage in a sensing cycle (i.e. $D_{t,h}=0$), the accessibility index is increased by $\kappa$. Intuitively, a lower $X_{t,h}$ means that a cell has a higher chance of being damaged based on prior history and should be disregarded from routing decisions. On the other hand, a higher $X_{t,h}$ indicates that a cell has less chance of damage based on historical damage information.

We employ an established graph traversal algorithm, the A* search algorithm~\cite{duchovn2014path}, to direct the scout cars in traversing the path covering the highest number of non-repeating edges between pairs of farthest nodes. The accessibility index is used by the Vehicle Dispatch (VD) module discussed in Section 4.3.2 to decide the route selection strategy for cars involved in event exploration.

\subsection{Vehicle Dispatch (VD) Module}
The VD module is designed to take the filtered social signals from the SSD module and the road damage information in the form of accessibility index from the RDD module for appropriately dispatching a group of interested vehicles to probable event locations. In particular, we use a Bottom-Up Game-Theoretic (BGT) policy to prioritize and allocate tasks to the cars based on the assigned task rewards, the distance between the cars and the event locations, the remaining time of the tasks, and the event uncertainty. Once the allocation of the tasks to the cars is completed by the BGT module, a Markov Decision Process (MDP)-based approach is employed to select the best available routing strategy for the cars while incorporating the road damage information. 

\subsubsection{Bottom-Up Game-Theoretic Task Allocation}
The bottom-up game-theoretic (BGT) task allocation approach is designed to allow the car drivers to make choices of event locations to travel to. The key motivation behind this design principle of the VD module is to let the car drivers express their individual task preferences in the allocation process. This allows the cars to determine the strategy that maximizes their individual payoffs~\cite{rashid2020socialdrone}. 

In game theory, congestion games are typically used to mitigate resource conflicts (e.g., event locations) among a set of players (e.g., cars). We adopt singleton weighted congestion games \cite{ieong2005fast}, a variant of congestion games where the expected utility of each task uniformly decreases as the  sum of players (cars) that picked the task increases. 
Moreover, each car only picks one task at a time according to the singleton property. 
The Pure Strategy Nash Equilibrium is guaranteed to exist under the above singleton weighted congestion game protocol~\cite{rashid2019socialcar}. This property enables the cars to make conclusive task allocation decisions.  In particular, where are four core components in our singleton weighted congestion game protocol: the \emph{reputation}, the \emph{reward}, the \emph{weighted congestion rate}, and the \emph{utility function}. We elaborate on them below.

Similar to the damage discovery module, we assume that a task can be picked up by multiple cars and assign a \textit{reward} $r_{t,n}$ for each task to incentivize the drivers to pick it up. We maintain a \textit{reputation score} $\pi_{t,p}$ for each car $C_p$ based on the historical performance till the response cycle $t$. We use $\nu_{t,p}$ to count the number of tasks successfully completed by car $C_p$, and $\tau_{t,p}$ to count the tasks marked as unsuccessful  (i.e. a car not being able to perform a task by the sensing deadline) up to response cycle $t$, respectively. Intuitively, if a car picks up a task and successfully completes it, the reputation score will increase. If the car fails to reach the destination on time or drops the task, the score will decrease accordingly. The \textit{reputation score} $\pi_{t,p}$ is based on an initial reputation $\pi_{0,p}$ at $t=0$ and is subsequently computed as:
\begin{equation}\label{eq:reputation}
    \pi_{t,p}=\pi_{t-1,p} + \eta \times (\sum{\nu_{t,p}}-\sum{\tau_{t,p}}), t>0
\end{equation}
where $\eta$ is an adjustable parameter called \textit{reputation coefficient}. If $\eta$ is set  high, the \textit{reputation score} will be more sensitive to the success and failure in the completion of tasks.  

We define a key component of our congestion game called \textit{weighted congestion rate} as follows: 
\begin{myDef}
	\emph{\textbf{Weighted Congestion Rate $\gamma^m_{t,n}$ for task $V_{t,n}$ for car $C_m$:} A score in the range of (0, $\infty$) that indicates the level of contention on a task. It serves as a discounting factor of the utility function to dissuade cars to pick the same task already selected by several cars. The weighted congestion rate is computed by:
	}
	\vspace{-0.1in}
\end{myDef}
\begin{equation}\label{eq:congestion}
\begin{split}
    \gamma^m_{t,n} = \sum_{\substack{p=1\\p\neq m}}^G S \times (\pi_{t,m}-\pi_{t,p})^k
\\S=
\begin{cases}
sgn(\pi_{t,m}-\pi_{t,p}), & \text{\textit{k} is even}\\
1, &\text{otherwise}\\
\end{cases}
\end{split}
\end{equation}
where $k$ is an exponential scaling factor to adjust the intensity of the congestion property. If $k$ is set to be high, the congestion rate will be more sensitive to the difference in the reputation scores. The intuition here is that if several cars with reputation scores greater than car $C_m$'s reputation score have already picked up a particular task, the congestion is higher. On the contrary, if a few cars have already picked the event and have lower reputation scores, the congestion will be lower.

We anticipate that once all the cars select all the tasks, a churn situation can occur. For example, a car may drop a task at any instant abruptly, new cars may join or existing cars may leave the system. This may necessitate a reallocation of the tasks. 
We keep track of the \textit{remaining time} $\rho_{t,n}$ for each task at any time instant and define it as:
\begin{equation}\label{eq:remaining}
\rho_{t,n}=\delta_{t,n}-\tau_t
\end{equation}
\noindent
where $\tau_t$ is the elapsed time from the beginning of the response cycle $t$.

Given the definitions above, we can now derive the \textit{utility function} based on which  the cars decide their best strategies and define it as:
\begin{myDef}
	\emph{\textbf{Utility Function $u^m_{t,n}$ for task $V_{t,n}$:} the utility function represents the benefit for picking a specific task (i.e., event location) for car $C_m$. 
	}
\end{myDef}

In our model, we devised a customized  utility function for car $C_m$, referred to as \textit{event priority score} as follows:
\begin{equation}\label{eq:priority}
u^m_{t,n} = 
\begin{cases}
\frac{r_{t,n} \times (\lambda_1 \times \omega^m_{t,n}+\lambda_2 \times \rho_{t,n}+\lambda_3 \times h(\Lambda_{t,n}))} {\gamma^m_{t,n}}, &\rho_{t,n}>0 \\
0, &\rho_{t,n}=0\\
\end{cases}
\end{equation}

The above utility function prioritizes the tasks for car task allocation based on four factors: i) the reward for the task, $r_{t,n}$; ii) the distance from the car to the event location, denoted as $\omega^m_{t,n}$; iii) the remaining time of the task, $\rho_{t,n}$; and iv) the uncertainty of an event, as captured by a function of the estimation confidence score (i.e. $f(EC_{t,n})$) from Definition \ref{confidence}. Given the rewards for the tasks, each car tries to prioritize the tasks with higher rewards. In particular, the remaining time factor prioritizes the tasks with tighter remaining deadlines while the distance factor priorities tasks with shorter distances from the cars in order to reach nearby tasks first. $\lambda_1$, $\lambda_2$, and $\lambda_3$ represent the weights of each factor. Their values are computed using \textit{proportional control}, a widely used control technique \cite{doyle2013feedback}. In Section \ref{sec:eval}, we discuss how the three parameters are determined. Finally, the congestion rate, $\gamma^m_{t,n}$ on the denominator of the utility function is designed to avoid contention of cars for a task. We highlight that multiple cars have the freedom to select the same task after all the tasks are allocated to at least one car. This is to increase the value of the congestion rate, thereby reducing the utility for each car. However, this approach also increases the chance of a task being completed. Additionally, if the remaining time $\rho_{t,n}$ is $0$, the utility is $0$.

Afterward, each car  decides on its best strategy towards maximizing its utility in the congestion game, until a Nash Equilibrium is reached. The Nash Equilibrium (NE) exists in the proposed game where each car is assumed to have determined its optimal decision (i.e., picking the task has the highest utility) and no car has anything to gain by only changing its preferred tasks.  We exploit the \textit{best-response dynamics} algorithm to find the NE~\cite{rashid2020compdrone}. 

We use an array $U_{t,n}$ for each task $V_{t,n}$ to record all the cars that pick the task in the $t^{th}$ response cycle after the NE is reached. Once the BGT sub-component determines the \textit{destinations} for the cars, the MDP scheme incorporates the road damage information from the RDD module to assign the best routes to destinations for the cars.

\subsubsection{Markov Decision Process (MDP)-based Route Selection Strategy}
As the accessibility index (indicating the possibility of road damage) is obtained from the RDD module and the destinations for the cars are derived from the BGT sub-component, the knowledge is utilized to perform routing decisions. We found that our problem of allocating traversable routes for cars nicely fits into the principle of Online Markov Decision Process (MDP)~\cite{even2009online}. For our problem, we consider the starting location of each car at every response cycle as a \textit{source} and the event location assigned to the car as a \textit{destination}. We assume that most pairs of source-destination are connected by multiple routes. Based on this assumption, we consider the source-destination pairs as the \textit{states} for our MDP model. In our model, we map the \textit{actions} to the list of available routes for each state and the \textit{penalties} (i.e., or equivalently negative reward) to the sum of the road damage for each action (i.e., route). Afterward, we develop a custom action selection scheme to determine the best actions and solve our MDP problem. Our choice for developing our own approach for solving the MDP problem is driven by the rationale that our environment is highly uncertain due to the dynamics of the social media and the physical world, which makes the determination of the best actions challenging. While our states and actions do not change, the values of the penalties (i.e., road damages) for corresponding actions often exhibit a dynamic behavior across response cycle due to the constantly changing number of event reports in the social media, their locations in the real world, and the road damage situation along the routes. This dynamism makes the determination of the best actions challenging. Moreover, while the scout cars provide the framework with limited information related to the road damages in a prior response cycle, we may not have the complete information of all the damaged roads across all the routes with a limited number of scout cars.

Our scheme for determining the best actions is built on a feedback control mechanism that uses the penalties as a feedback signal. At the end of every response cycle, the penalties from the current sensing cycle are updated based on Equation~\ref{eq:penalty} which is a function of the predicted and actual road damage for the current sensing cycle. \mt{The actions to take in the next sensing cycle are determined by Equation~\ref{eq:actionprob} which is a function of the prior road damage and the normalized difference between the sums of the penalties across successive response cycles.} The details of the states, actions, and rewards, as well as the mechanism of our algorithm for selecting the best actions, are discussed below.

\begin{myDef}
\emph{\textbf{States $\mathcal{W}^t$:} A set of tuples $\mathcal{W}^t=\{W^t_1,W^t_2,...,W^t_{j^t}\}$ denoting source-destination pairs at the $t^{th}$ response cycle. An example of a source-destination pair in response cycle $t=3$ is $W^3_1=(SC_{3,2},SC_{3,8})$, where $SC_{3,2}$ is the source sensing cell (i.e., the position of the car) and $SC_{3,8}$ is the destination sensing cell (i.e., the event location).
}
\end{myDef}

\begin{myDef}
\emph{\textbf{Actions $\mathcal{A}_k^t$ for state $W^t_k$:} A set of ordered lists $\mathcal{A}_k^t=\{A_{k,1}^t,A_{k,2}^t,...,$ $A_{k,{l^t}}^t\}$ representing all the available routes for each state $k$ at sensing cycle $t$. Each action set $\mathcal{A}_k^t$, consisting of a set of contiguous sensing cells, maps to each state $W_k^t$. For example, actions set $\mathcal{A}_1^3$ for state $W^3_1$ at $t=3$ can have an action $A_{1,4}^3$ with sensing cells $[SC_{3,2},$ $SC_{3,6},$ $SC_{3,8}]$.}
\end{myDef}

All the available routes for each state are obtained by considering the task deadlines and sensing cell constraints for the cars while reaching the assigned sensing cells. We leverage a route planning algorithm based on a Contraction Hierarchies technique from graph theory~\cite{geisberger2008contraction} to generate all the available actions for each state.

The probability function for the $v^{th}$ action, $A_{k,v}^t$ in an action set is given by:
\begin{equation}\label{eq:actionprob}
P(A_{k,v}^t)=\sigma_{t-1}\prod_{h\in a_{k,v}^t} X_{t-1,h}
\end{equation}
where $\sigma_{t-1}$ is a parameter called \textit{penalty differential} that is determined by the penalty from the prior response cycle in Equation~\ref{eq:penaltydiff} discussed later. The term $a_{k,v}^t$ represents the set of sensing cells that are part of the route for action $A_{k,v}^t$.

At the beginning of every sensing cycle $t$, the accessibility index indicating the possibility of road damage from the last sensing cycle (i.e. $t-1$) is used in Equation \ref{eq:actionprob} to generate the probability of all the actions that can be taken in the current sensing cycle (i.e. $t$). Intuitively, the greater the accessibility indices for the associated cells, the higher the probability for that action consisting of the particular cells to be taken.

We anticipate that it might not possible to locate all the road damage by the scout cars in the RDD module. Moreover, the road damage could be encountered later by the other cars assigned for event exploration.
\begin{myDef}
\emph{\textbf{Penalties $\mathcal{R}^t$:} A set of penalty scores $\mathcal{R}^t=\{R_{1}^t,R_{2}^t,$ $..., R_{j^t}^t\}$ obtained by summing the road damages discovered by the car drivers tasked with event exploration for each action that is selected for each state. Each element of the set represents a particular action selected for the corresponding state and is thereby computed by:}
\begin{equation}\label{eq:penalty}
R_{u}^t=\sum_{h\in a_{u}^t} (\overline{D_{t,h}} \lor D_{t,h})
\end{equation}
\end{myDef}
where $\overline{D_{t,h}}$ represents the road damage discovered by the cars exploring events at sensing cell $SC_{th}$ and $a_{u}^t$ represents the set of all the sensing cells in the selected action.  Initially, $\overline{D_{t,h}}=D_{t-1,h}$ for all the cells at the beginning of every sensing cycle. Once an event exploration car discovers damage in a sensing cycle, $\overline{D_{t,h}}=1$. In a future sensing cycle if the damage appears to be repaired, $\overline{D_{t,h}}=0$. Intuitively, if $R_l=0$, the route is assumed to be fully accessible. Otherwise if $R_l>0$, the route may contain damage and should be avoided. We adjust the penalty differential $\sigma_{t}$ discussed earlier based on the normalized difference between the sums of the penalties across successive response cycle:
\begin{equation}\label{eq:penaltydiff}
\sigma_{t}=\sigma_{t-1}-\frac{\sum\mathcal{R}^t-\sum\mathcal{R}^{t-1}}{\sum\mathcal{R}^t+\sum\mathcal{R}^{t-1}}
\end{equation}
The rationale is that if the total magnitude of discovered damage increases in the current response cycle, we lower the penalty differential to ``penalize" the action taken earlier.

We formally define our MDP model below. We consider an MDP with a state set $\mathcal{W}^t$ mapped to the source-destination pairs, an action set $\mathcal{A}_k^t$ mapped to the choice of routes, and a penalty set $\mathcal{R}^t$ mapped to the aggregate road damages discovered by all the car drivers for the selected actions. The goal of the MDP is to derive an optimal collection of actions that minimizes the encounter of road damages by the cars assigned for event exploration. Formally the objective is:
\begin{equation} \label{eq:objMDP}
\begin{split}
& \operatorname*{argmin}_{\mathcal{A}_k^t} \sum_{t=1}^{T} \mathcal{R}^t, 1  \le t \le T\\
\end{split} 
\end{equation} 
\noindent
The optimal action set for a corresponding state can be obtained by incorporating the \textit{contextual epsilon-greedy strategy} \cite{raykar2014sequential}. The first step is to initiate a learning phase (i.e., \textit{exploration}), for a certain duration of response cycles to determine the optimal actions $\mathcal{A}_k^t$ for each state. During the learning phase, the state is determined based on the source-destination pairs of each car and the action sets are generated by enumerating through all the possible routes. The scout cars are dispatched by the RDD module which continuously feeds the damage information to the VD module. An action $A_k^t\in\mathcal{A}_k^t$ is selected for each corresponding state $W_k^t$ that has not been previously explored. Cars are then dispatched for exploration using the selected action based on the sampling probability and the penalty $\mathcal{R}^t$ is observed. If the observed penalty changes, the penalty differential $\sigma_{t}$ is adjusted. Once the learning phase is complete, the probability of the actions in the action set $\mathcal{A}_k^t$ is updated. Subsequent response cycles use the present information to make route selections (i.e., \textit{exploitation}).

\subsection{Dynamic Incentive Control (DIC) Module}
The Dynamic Incentive Control (DIC) module is incorporated to complement the VD module and mitigate the churn issue that may prevail in the system. We utilize a top-down optimal control to adjust the rewards for the events based on the attribute of the cars $U_{t,n}$ for selecting the tasks at every response cycle. Intuitively, if a lot of tasks are being dropped, we may want to increase the rewards for the particular tasks to encourage more drivers to pick up those tasks.

\subsubsection{Top-Down Optimal PID Controller}
The reward for each task is assigned based on an \textit{initial reward} $r_0$ and a \textit{reward adjustment function} $q_{t,n}$ as expressed below:
\begin{equation}
    r_{t,n}=r_0+q_{t,n}
\end{equation}

A na\"ive solution to decide the value of $q_{t,n}$ would be to set it proportional to the number of dropped tasks after each response cycle $t$. However, this approach may not be optimal as it would infrequently set rewards and make the system unstable (i.e. the rewards may fluctuate uncontrollably if too many tasks get dropped). To address this problem, we incorporate a proportional-integral-derivative (PID) controller, a robust control loop feedback mechanism used in industrial control systems as well as applications requiring continuously modulated control. The PID controller nicely maps to our problem of determining the value of $q_{t,n}$. The process variable in the PID controller is the aggregate reputation of the cars that pick a particular task which is formally defined as:

\begin{myDef}
\emph{\textbf{Aggregate Reputation $e_{t,n}$ for task $V_{t,n}$:}
The sum of the reputations of all the cars that selected task $V_{t,n}$.
\begin{equation}
    e_{t,n}=\sum_{p\in{U_{t,n}}} \pi_{t,p}
\end{equation}
Intuitively, the higher the value of $e_{t,n}$, the greater the chance for the cars to make successful attempts to complete the task. 
}
\end{myDef}
We consider that the framework has a settable parameter called base reputation score $e'$, which defines the worst-case aggregate reputation for all the tasks acceptable by the system at any response cycle. If the aggregate reputation falls below this threshold for any task, the system aims to recover the performance by increasing the rewards assigned for the specific task. On the other hand, if the score is above the threshold, the system makes a decision to lower the reward for the particular task, and the surplus could be allocated elsewhere with other tasks. We map the base reputation score $e'$ as the set  point for the PID controller and the aggregate reputation score $e_{t,n}$ as the measured process variable. Thus, the error for the PID controller is given by:
\begin{equation}
    \overline{e_{t,n}}=e'-e_{t,n}
\end{equation}


The system constantly monitors the number of tasks that are dropped by the cars (i.e. churn). Every time the system observes that the number of dropped tasks exceeds a certain threshold $\psi$, it reruns the algorithm for computing the rewards for the tasks and the utilities for the cars. Otherwise, the algorithm is run periodically at every response cycle for allocating the tasks and to cater to updated event reports. 

	\section{Evaluation} \label{sec:eval}
In this section, we evaluate the performance of DASC through a real-world post-disaster case study involving road damage scenarios. The evaluation results exhibit significant performance gains of DASC over the compared baselines in terms of both detection effectiveness and deadline hit rate in verifying the disaster events while considering the road damage.

\subsection{Experimental Setup}
\begin{wrapfigure}{l}{0.40\textwidth}
    \centering
    \includegraphics[width=6cm]{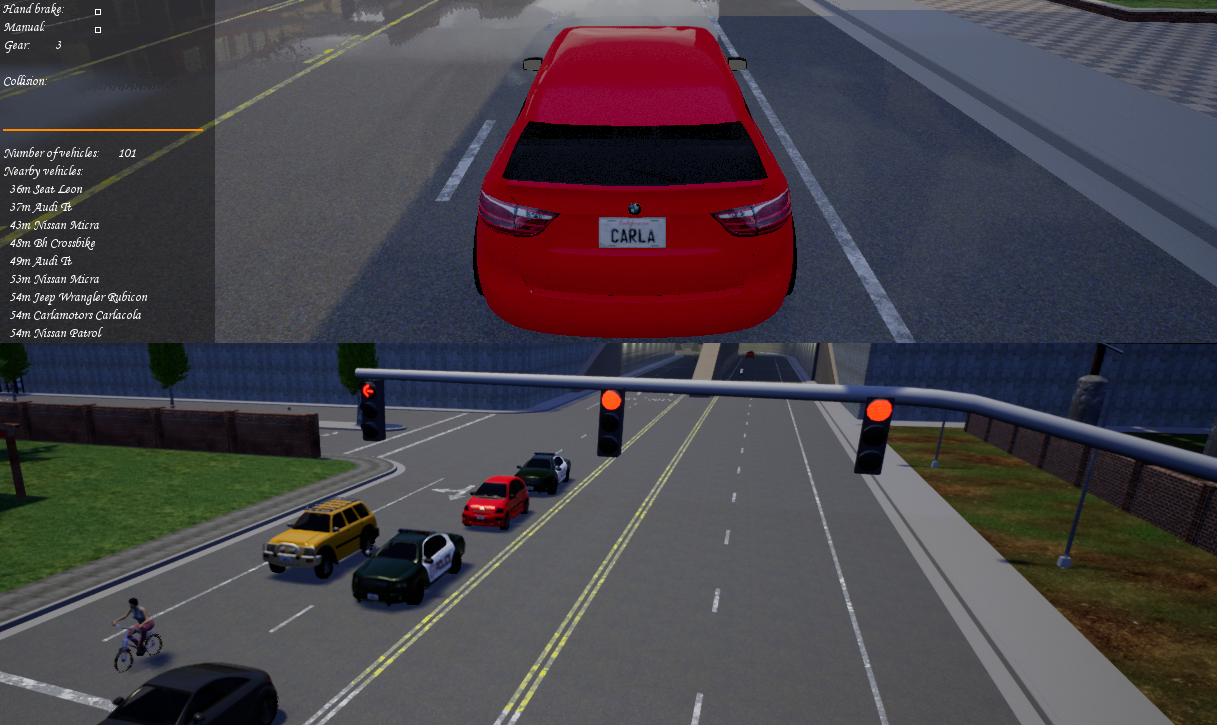}
    \vspace{-0.1in}
    \caption{CARLA interfaced with DASC. The top pane displays the third-person view of a single vehicle while the bottom pane shows the view of multiple vehicles.} 
    \label{fig:CARLA}
\end{wrapfigure}

We acknowledge the fact that the deployment of vehicles in an actual disaster scenario is either impossible or immensely difficult because a real-world disaster is hard to predict and cannot be reproduced. As such, we carry out a real-world data-driven emulation to evaluate our system. The evaluation platform consists of three key components: 1) the CARLA simulator; 2) a real-world mapping interface; and 3) the DASC system. CARLA is a widely used car simulator that can closely imitate physical models of cars traveling in the physical world along with congestion and traffic signals at intersections~\cite{dosovitskiy2017carla}. Figure \ref{fig:CARLA} shows a snapshot of our emulation environment. 

The real-world mapping interface integrates the CARLA simulator with OpenStreetMaps~\cite{haklay2008openstreetmap} to replicate the real-world map. This enables CARLA to simulate dispatching cars to real-world locations (i.e. addresses reported in the social media). The DASC scheme generates the tasks and rewards for dispatching the cars along with their corresponding routes. The DASC scheme connects with the CARLA simulator using a Python API~\cite{dosovitskiy2017carla} to send commands for simulating the actual car route in the real-world. Figure \ref{fig:CARLA} illustrates a snapshot of the CARLA simulator interfaced with the DASC framework.
\subsection{Parameter Tuning}
	In order to obtain the optimized values of the parameters $\lambda_1$, $\lambda_2$, $\lambda_3$, $K_p$, $K_i$, $K_d$, $\psi$, and $\kappa_t$, we carry out a parameter tuning process in the first $1/4^{th}$ of the response cycles. We select the F1 score as the optimization objective as it can give a better measure of the incorrectly classified cases with imbalanced distributions~\cite{joshi2016accuracy} such as in our case with social media data. Since the input parameters cannot be directly modeled on the F1 score (i.e., using a mathematical equation), we incorporate a non-linear optimization~\cite{floudas1995nonlinear} approach for obtaining the values of the parameters. We first set the initial values of all the parameters to a maximum value of $1$. At each response cycle during the training, cars are dispatched using the BGT task allocation scheme. We locate the parameter values that yield the maximum F1 score by using the Nelder-Mead method~\cite{luersen2004globalized}. Specifically, we split the training phase into 3 equal segments and after a series of car observations are collected for every segment, the values of the parameters that increase the F1 score are retained. We then apply a non-linear optimization on the values of the parameters and assign the final values of the parameters. We determined the values of the parameters as: $\lambda_1=0.82$, $\lambda_2=0.58$, $\lambda_3=0.49$, $K_p=0.11$, $K_i=0.67$, $K_d=0.38$, $\psi=0.62$, and $\kappa_t=0.65$. \mt{For the initial accessibility index $X_{0,h}$, our objective is to determine a value that minimizes the difference between the accessibility index $X_{t,h}$ and the actual road damage $D_{t,h}$.} Therefore, instead of the complex Nelder-Mead method, we use a linear time optimization~\cite{pilanci2017newton} approach to determine $X_{0,h}$, which is found to operate optimally approximately at the mid-range of $X_{t,h}$ (i.e. $X_{0,h}=0.5$). After obtaining all the parameters, the system is expected to work in its optimal state by determining the best task choices and rewards.

\subsection{Evaluation Dataset}
We collected a real-world dataset using Twitter data feeds posted immediately after the 2017 Hurricane Harvey, a hurricane marked as the costliest tropical cyclone, causing \$125 billion in damage. The hurricane originated from a destructive rainfall-triggered flood in the Houston metropolitan area and Southeast Texas in August 2017~\footnote{https://www.ahcusa.org/harvey-8252017.html}. To obtain the road damage information of the disaster, we analyzed the 2017 Hurricane Harvey damage report map published by FEMA and deduced the roads affected by damage during the disaster~\cite{murphy2017harvey}. We then replicated the road damage in the CARLA simulator.

We obtained the Twitter data using the Apollo data collection tool~\footnote{http://apollo.cse.nd.edu/index.html}. For the evaluation purpose, we only consider the data related to critical resources (i.e., gas station and pharmacy availability). We separately gathered the ground truth labels of the reported events from historical facts published by credible sources (i.e., disaster reports) such as \cite{moran2015speeding}.  The statistics of the dataset are summarized in Table \ref{tab:stats}.

\begin{wraptable}{l}{0.40\textwidth}
\scriptsize
\vspace{-0.2in}
    \caption{Data Statistics}
    \centering
    \begin{tabular}{|l|l|}
       \hline
       Start Date & August 27, 2017\\ \hline
        Time Duration & 3 days \\ \hline
       Location & Houston, Texas, USA\\ \hline
        No. of tweets & 1,691 \\ \hline
        No. of tweet users & 1,446 \\ \hline
       No. of event locations & 106 \\ \hline
   \end{tabular}
    \label{tab:stats}
\end{wraptable}

The social sensing component for our framework as well as the baselines use the Robust Truth Discovery (RTD)~\cite{zhang2016robust} algorithm for deciding whether an event occurs or not. We replay the obtained data trace to emulate the disaster event. We sort all the reported events based on their timestamps and distribute them across different response cycles. For our particular experiment, we selected the duration of each response cycle to be 100 minutes based on the frequency of the events observed in our dataset. There are a total of 36 response cycles. Within each response cycle, a set of data preprocessing steps are performed. In particular, we extract the relevant tweets by first running keyword searches (e.g., gas, fuel, oil, medicine, healthcare, and pharmacy) and discard the irrelevant ones. We then cluster similar tweets into the same groups using the state-of-the-art online tweet clustering tool~\cite{zhang2016robust} and obtain claims that report events at particular locations. Also, we only keep the tweets that have valid geo-location tags for our experiments. 

\subsection{Compared Baselines}
We compare the performance of DASC with a few representative baselines. We first acknowledge the fact that we have not come across any solution that guides vehicles for sensing using social media signals and simultaneously incorporates the dynamics of the social media (i.e., evolving number of events and social media users), the physical world (i.e., the road damage and the deadline of the events), and the rationale behavior of the car drivers (i.e., churn). Therefore, we included five established VSN-based event discovery schemes from current literature. Since the schemes do not incorporate any social sensing component, we included our previous SocialCar framework as a baseline as well a simplified version of the DASC framework called ``DASC w/o MDP" to demonstrate the impact of the social signals.

\begin{itemize}
    \item \textbf{Random Allocation}:  tasks are allotted randomly to cars on the roads. Once cars come across an event, they record and report it. Commercial crowdsourcing platforms like Waze \cite{galeso2016waze} use this technique.
    \item \textbf{Fixed Route}:  a fixed number of dedicated cars traverse along designated patrol routes. 
    A patrol route is designed by covering the maximum number of sensing cells using a Hamiltonian Cycle-based approach~\cite{portugal2010msp}.
    \item \textbf{Shortest Distance Based}:  cars that are in closer proximity to event locations are prioritized first with the assumption that they have higher chances of reaching the destinations faster~\cite{palazzi2012delay}. 
    \item \textbf{Reputation Based}:  tasks with the shortest deadlines are assigned to cars with the highest reputation first~\cite{gong2014social}. 
    \item \textbf{Incentive Based}:  tasks with the shortest deadline get the highest rewards in the task allocation process~\cite{gao2015survey}.
    \item \textbf{SocialCar}:  a simplified version of the DASC scheme that assigns tasks to cars solely based on the social media reports and the reputation of cars, and adjusts the incentives based on our previous work~\cite{rashid2019socialcar}. The SocialCar scheme does not consider the road damage along any route.
    \item \textbf{DASC w/o MDP}:  a simplified version of the DASC scheme without the MDP component. Once road damage is discovered, the cars are na\"ively assigned the routes with the highest accessibility indices.
\end{itemize}

\subsection{Evaluation Results}
We conduct four different sets of experiments to extensively assess the performance of all the schemes using the real-world dataset. We considered three types of drivers in our evaluation: i) drivers who accept tasks and attempt to successfully complete the tasks; ii) drivers who accept tasks and randomly abort midway; iii) drivers who are unwilling to participate in the sensing application. We maintain an equal proportion of cars across all three categories in the first three sets of experiments to observe the impact of other variables. However, in the fourth experiment, we analyze the effect of varying the ratios of the three types of drivers on the performance of all schemes.

\subsubsection{Detection Effectiveness}
In the first set of experiments, we assess the performance of all schemes across the entire dataset. The detection effectiveness is evaluated using common metrics for binary classification: \emph{Accuracy}, \emph{Precision}, \emph{Recall}, and \emph{F1-Score}. We utilized a set of 90 cars in our system. The results are presented in Table \ref{tab:HarveyOverall}. We discover that DASC outperforms the other schemes in identifying the truthful events (i.e., gas station and pharmacy availability) in the aftermath of Hurricane Harvey. In terms of classification accuracy, precision, recall, and F1 score, the performance gains achieved by DASC compared to the best-performing baseline (i.e., the \textit{SocialCar} scheme) are 4.4\%, 13.4\%, 1\%, and 9.5\%, respectively. Such increased performance highlights the importance of incorporating the top-down incentive control in the task allocation process. Since the rewards are adjusted dynamically based on the reputation of the cars that select the events, the system maximizes the possibility of completing the tasks. In addition, we observe that DASC outperforms other baselines by a fairly large margin. We accredit this performance gain to the design of DASC that explicitly considers the road damages along the cars' routes and seamlessly integrates the social sensing and vehicular sensing system.

\begin{table}[htb!]
  \centering
  \caption{Overall Performance with Hurricane Harvey Dataset}
  \scalebox{0.89}{
	\begin{tabular}{c|| c c c c }
\toprule
Algorithm & Accuracy & Precision & Recall & F1-Score\\

\cmidrule(l){1-5}
    Random Allocation  & 0.111 & 0.131 & 0.323 & 0.186\\
    Fixed Route  & 0.239 & 0.267 & 0.517 & 0.352\\
    Shortest Distance  & 0.375 & 0.483 & 0.581 & 0.528\\
    Reputation Based     & 0.343 & 0.423 & 0.582 & 0.490\\
    Incentive Based     & 0.338 & 0.445 & 0.568 & 0.498\\
    SocialCar     & 0.613 & 0.527 & 0.818 & 0.641\\
    DASC w/o MDP & 0.430 & 0.502 & 0.668 & 0.573\\
\cmidrule(l){1-5}

\textbf{DASC} &\textbf{0.657} & \textbf{0.661} & \textbf{0.828} & \textbf{0.736}\\
\midrule
\toprule
    \end{tabular}
    }

\label{tab:HarveyOverall}
\end{table}

We further split the dataset across two different categories: i) gas station availability and ii) pharmacy availability in the Houston region for a more fine-grained assessment of all compared schemes. Table \ref{tab:Harvey1} shows the results. We observe that DASC continues to outperform all baselines across the split dataset.

\begin{table}[htb!]
  \centering
  \caption{Performance with Hurricane Harvey Dataset Across Different Categories}
   \scalebox{0.89}{
	\begin{tabular}{c|| c c c c || c c c c}
\toprule
& \multicolumn{4}{c}{Gas Station Availability} & \multicolumn{4}{c}{Pharmacy Availability}\\
\cmidrule(l){2-9}
Algorithm & Accuracy & Precision & Recall & F1-Score & Accuracy & Precision & Recall & F1-Score\\

\cmidrule(l){1-9}
   Random Allocation  & 0.126 & 0.148 & 0.334 & 0.205 & 0.093 & 0.110 & 0.303 & 0.162\\
Fixed Route &  0.244 & 0.272 & 0.519 & 0.357 & 0.234 & 0.261 & 0.513 & 0.346 \\
   Shortest Distance  & 0.402 & 0.507 & 0.598 & 0.549 & 0.352 & 0.463 & 0.569 & 0.511\\
   Reputation Based  & 0.362 & 0.437 & 0.592 & 0.503 & 0.327 & 0.413 & 0.576 & 0.481\\
  Incentive Based  & 0.355 & 0.476 & 0.566 & 0.517 & 0.315 & 0.407 & 0.561 & 0.472\\
  SocialCar  & 0.625 & 0.540 & 0.839 & 0.657 & 0.606 & 0.519 & 0.803 & 0.630\\
  DASC w/o MDP  & 0.435 & 0.499 & 0.685 & 0.577 & 0.415 & 0.494 & 0.635 & 0.555\\
\cmidrule(l){1-9}
\textbf{DASC} &\textbf{0.678} & \textbf{0.689} & \textbf{0.839} & \textbf{0.757} & \textbf{0.628} & \textbf{0.626} & \textbf{0.808} & \textbf{0.706}\\
\midrule
\toprule
    \end{tabular}
     }
\label{tab:Harvey1}
\end{table}


\subsubsection{Tuning the Number of Cars}
\begin{figure}[!h]
\vspace{-0.01in}
    \centering
    \includegraphics[width=12cm]{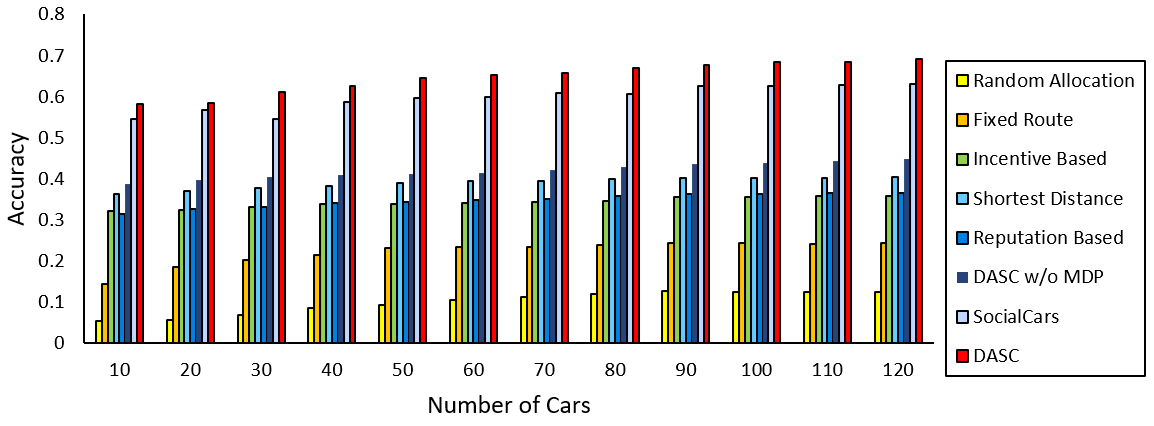}
    \vspace{-0.15in}
    \caption{Accuracy vs. Number of Cars}
    \label{fig:Accuracy}
    \vspace{-0.1in}
\end{figure}

\begin{figure}[!h]
\vspace{-0.01in}
    \centering
    \includegraphics[width=12cm]{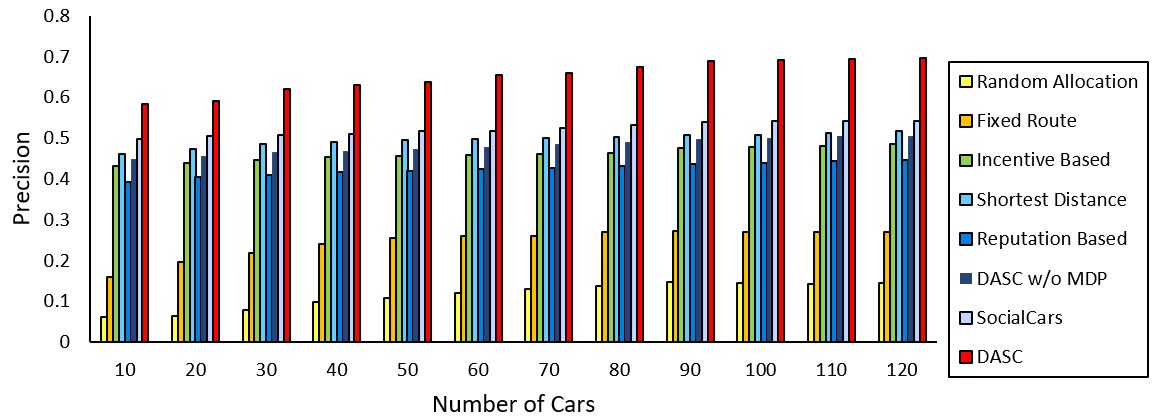}
    \vspace{-0.15in}
    \caption{Precision vs. Number of Cars}
    \label{fig:Precision}
    \vspace{-0.1in}
\end{figure}

\begin{figure}[!h]
\vspace{-0.01in}
    \centering
    \includegraphics[width=12cm]{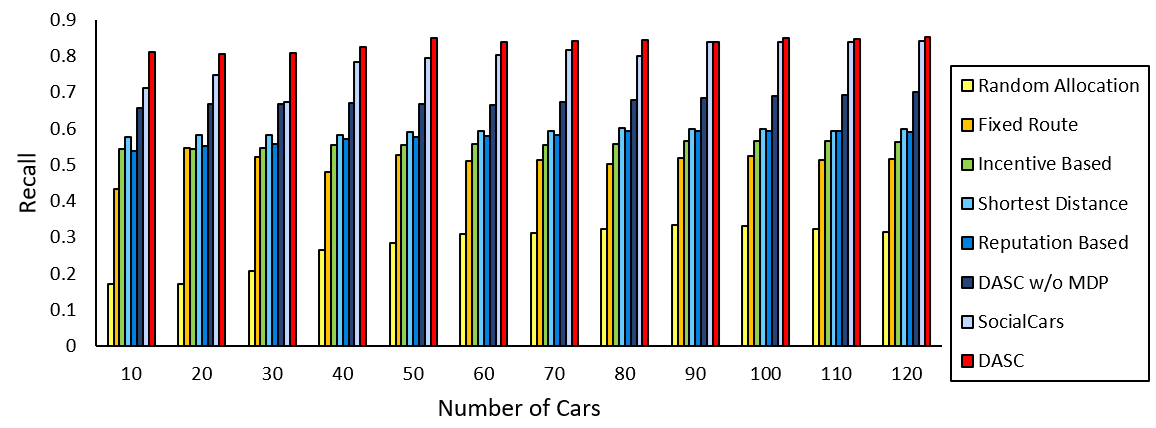}
    \vspace{-0.15in}
    \caption{Recall vs. Number of Cars}
    \label{fig:Recall}
    
    \vspace{-0.1in}
\end{figure}

\begin{figure}[!h]
\vspace{-0.01in}
    \centering
    \includegraphics[width=12cm]{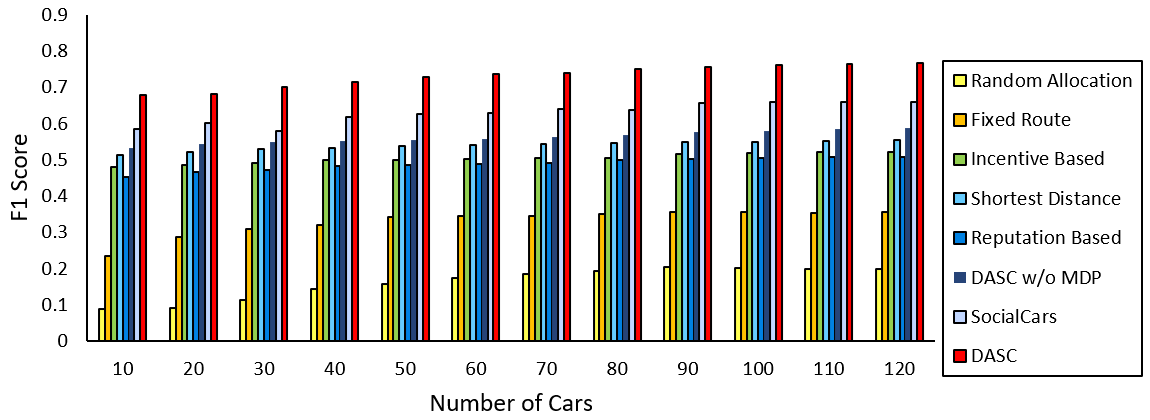}
    \vspace{-0.15in}
    \caption{F1 Score vs. Number of Cars}
    \label{fig:F1}
    \vspace{-0.1in}
\end{figure}

In the second set of experiments, we investigate the effect of the number of cars on the performance of all the tested schemes. For this assessment, we varied the number of cars across all the schemes within the City of Pasadena in the Houston Metropolitan area from our dataset. Figures~\ref{fig:Accuracy}, \ref{fig:Precision}, \ref{fig:Recall}, and \ref{fig:F1} show the results for accuracy, precision, recall, and F1 scores, respectively for all the compared schemes. We start with 10 cars and scale up gradually in increments of 10 cars for each round. We observe that the benefit obtained by increasing the number of cars starts to slowly plateau when the total number of cars reaches 100. We investigated this phenomenon and found two possible causes for this. Firstly, our dataset encompasses the relatively small city of Pasadena in Houston, Texas which is about 44.52 sq. mi. in size~\cite{pasadena}. Secondly, due to the hurricane, most road networks were rendered unusable leaving only a limited number of available routes for the cars. Given the small region and constrained road networks, increasing the number of cars for the sensing would not necessarily increase the performance. We observe that despite this, DASC manages to outperform all the baselines when changing the number of cars. This is because when churn issue occurs (i.e., cars drop tasks), the proposed reputation-based incentive adjustment, jointly with the deadline-aware bottom-up task allocation, allows the DASC to ensure the maximum number of events been covered in time. This eventually leads to better performance of DASC compared to other baseline schemes when the total number of cars are varied across the region. In addition to this, the damage-aware route allocation in DASC avoids routes that may have higher chances of road damage, thereby maintaining consistent performance.

\subsubsection{Deadline Hit Rate}
\begin{figure}[!htb]
    \centering
    \includegraphics[width=12cm]{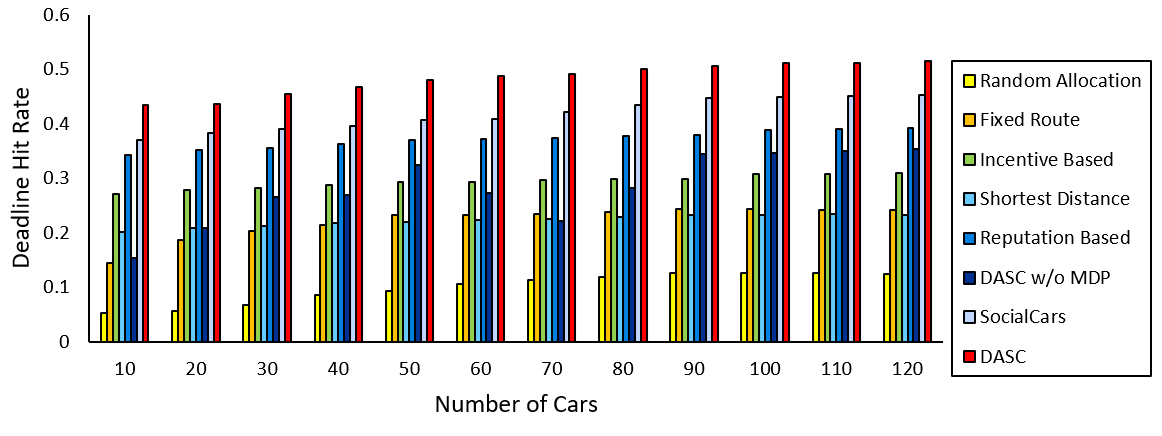}
    \vspace{-0.15in}
    \caption{Deadline Hit Rate vs. Number of Cars}
    \vspace{-0.3in}
   \label{fig:Deadlines}
\end{figure}

In the third set of experiments, we assess the deadline hit rate of all the compared schemes while varying the number of cars. Figure \ref{fig:Deadlines} shows the results. We observe that DASC achieves the highest deadline hit rate when the number of cars changes. Compared to the best-performing baseline, the SocialCar scheme, DASC achieves a 5.87\% better deadline hit rate with 90 cars. This is accredited to the combined effort of the bottom-up game-theoretic task allocation, the top-down incentive control, and the damage-aware route selection of the DASC system. At every response cycle, tasks having tighter deadlines are prioritized.

\subsubsection{Handling of Churn Issue}
In the fourth set of experiments, we study the \textit{churn} issue of the cars in the system and evaluate the robustness of the schemes against different behavior of the drivers. In our study, we found that it is very difficult to apply a concrete mathematical model to explicitly model the behavior of rational car drivers. The actions of car drivers are often unpredictable in real-world applications~\cite{wilde1976social}. In addition to that, during our findings, we did not come across any existing publicly available dataset that summarizes the behavior of the drivers who participate in roadside sensing. As such, it is impossible to predetermine what action each driver would take~\cite{wang2014modeling}. Thus, we design a set of simulation experiments to study the effect of driver behavior on the performance of DASC where we separately vary the proportion of car driver: i) who attempt to successfully complete the given tasks, ii) who abort tasks midway during exploration, and iii) who are unwilling to participate in the first place.

\begin{figure}[!h]
  \centering
  \begin{minipage}[b]{0.3\textwidth}
    \includegraphics[width=5.7cm]{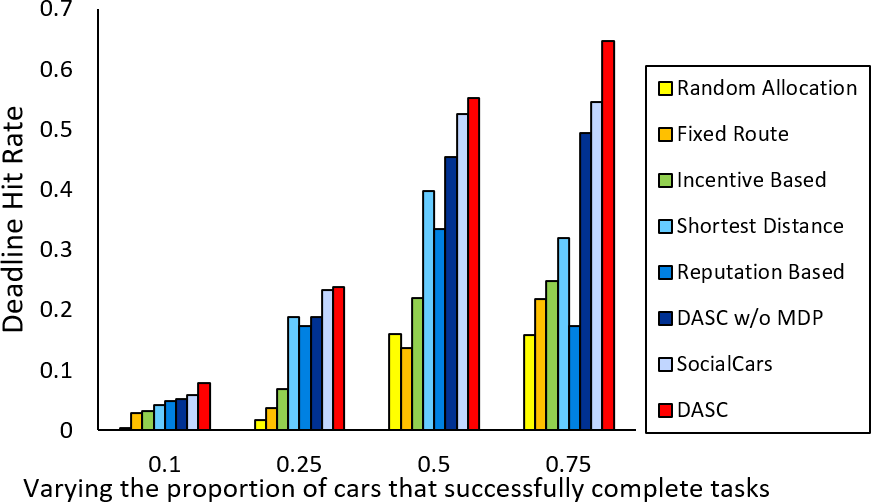}
    \vspace{-0.25in}
    \caption{Deadline Hit Rate vs. Proportion of car drivers that successfully complete tasks}
    \label{fig:Deadlines2}
  \end{minipage}
  \hfill
  \begin{minipage}[b]{0.3\textwidth}
    \includegraphics[width=5.7cm]{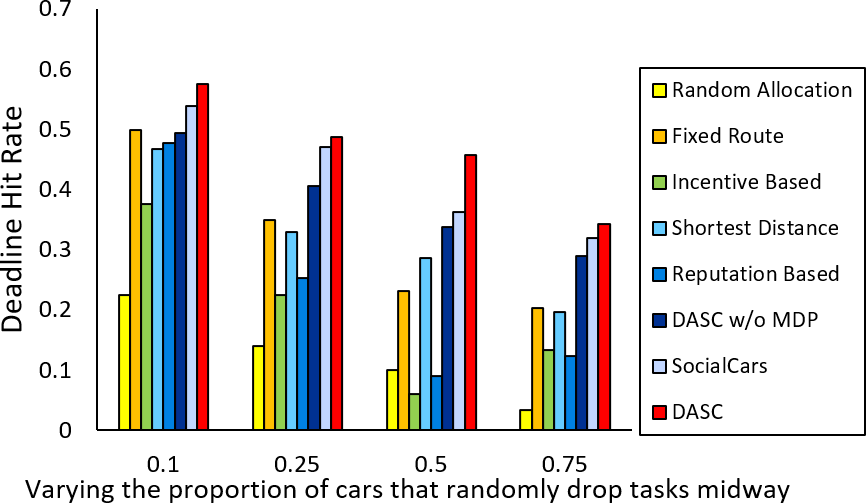}
    \vspace{-0.25in}
    \caption{Deadline Hit Rate vs. Proportion of car drivers that randomly drop tasks}
    \label{fig:Deadlines3}
  \end{minipage}
  \hfill
  \begin{minipage}[b]{0.3\textwidth}
    \includegraphics[width=5.7cm]{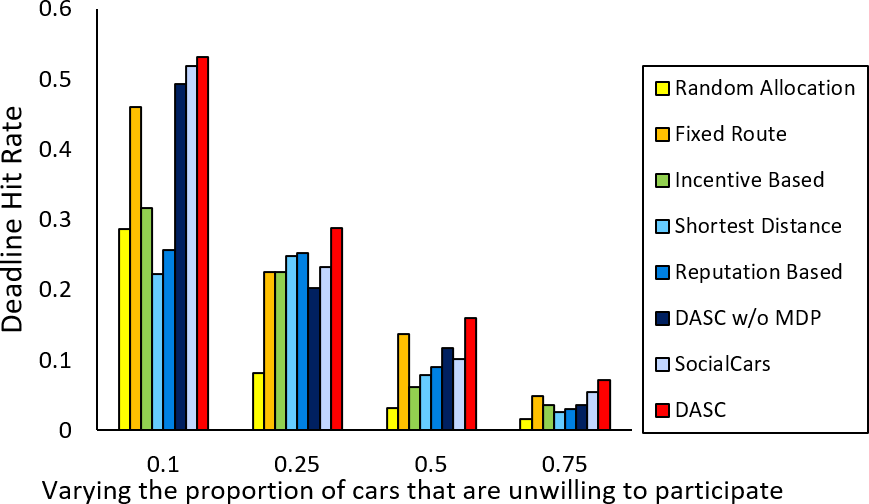}
    \vspace{-0.25in}
    \caption{Deadline Hit Rate vs. Proportion of car drivers that are unwilling to participate}
    \label{fig:Deadlines4}
  \end{minipage}
\end{figure}

\begin{figure}[!h]
  \centering
  \begin{minipage}[b]{0.3\textwidth}
    \includegraphics[width=5.7cm]{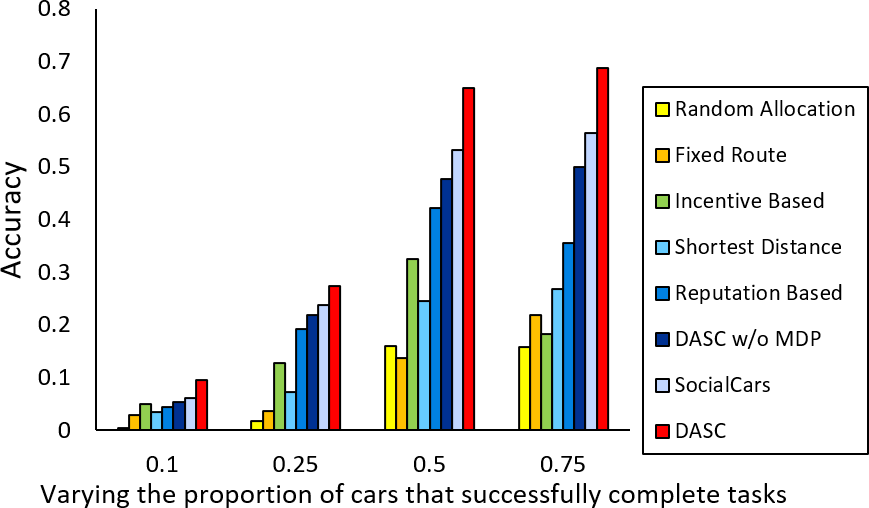}
    \vspace{-0.25in}
    \caption{Accuracy vs. Proportion of car drivers that successfully complete tasks}
    \label{fig:Accuracy1}
  \end{minipage}
  \hfill
  \begin{minipage}[b]{0.3\textwidth}
    \includegraphics[width=5.7cm]{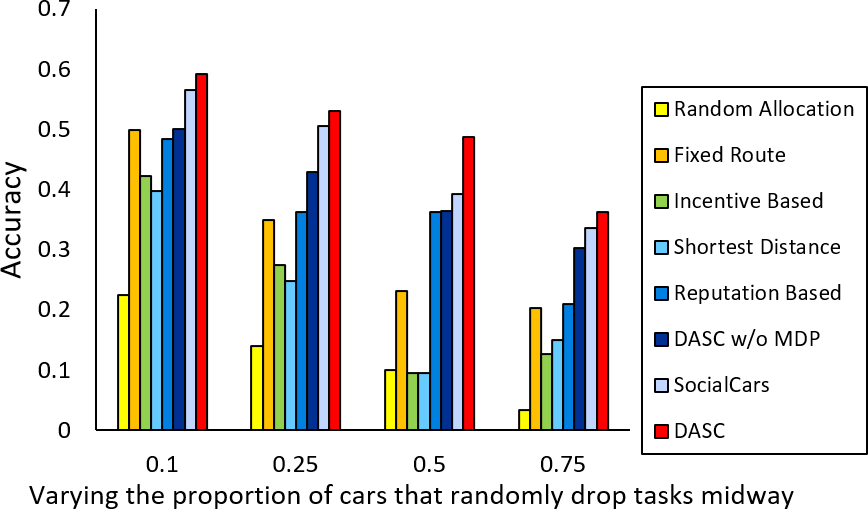}
    \vspace{-0.25in}
    \caption{Accuracy vs. Proportion of car drivers that randomly drop tasks}
    \label{fig:Accuracy2}
  \end{minipage}
  \hfill
  \begin{minipage}[b]{0.3\textwidth}
    \includegraphics[width=5.7cm]{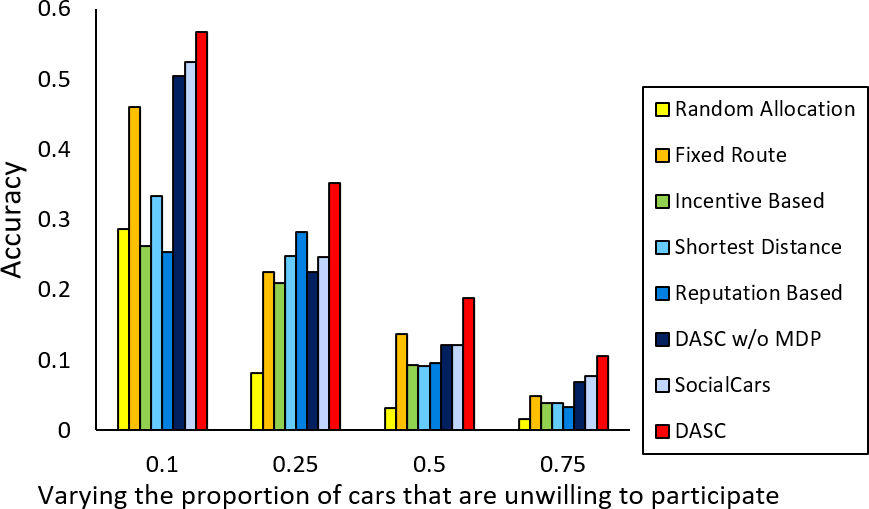}
    \vspace{-0.25in}
    \caption{Accuracy vs. Proportion of car drivers that are unwilling to participate}
    \label{fig:Accuracy3}
  \end{minipage}
\end{figure}

\begin{figure}[!h]
  \centering
  \begin{minipage}[b]{0.3\textwidth}
    \includegraphics[width=5.7cm]{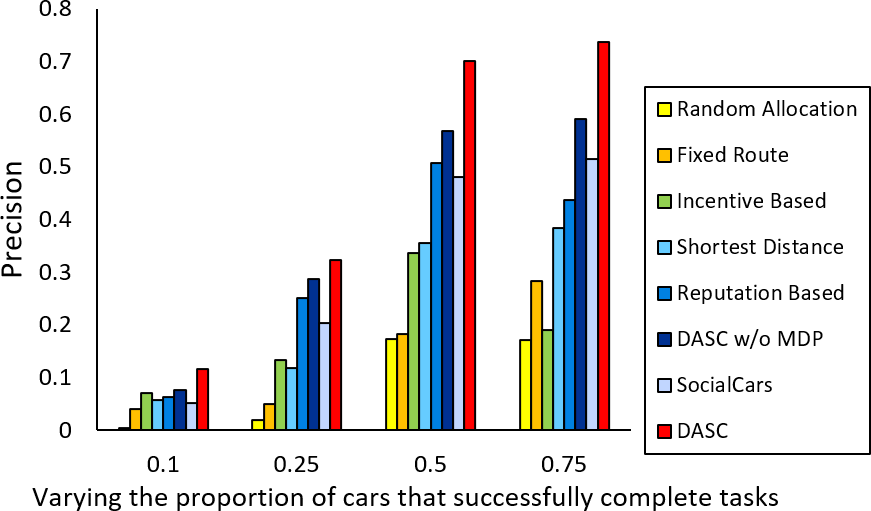}
    \vspace{-0.25in}
    \caption{Precision vs. Proportion of car drivers that successfully complete tasks}
    \label{fig:Precision1}
  \end{minipage}
  \hfill
  \begin{minipage}[b]{0.3\textwidth}
    \includegraphics[width=5.7cm]{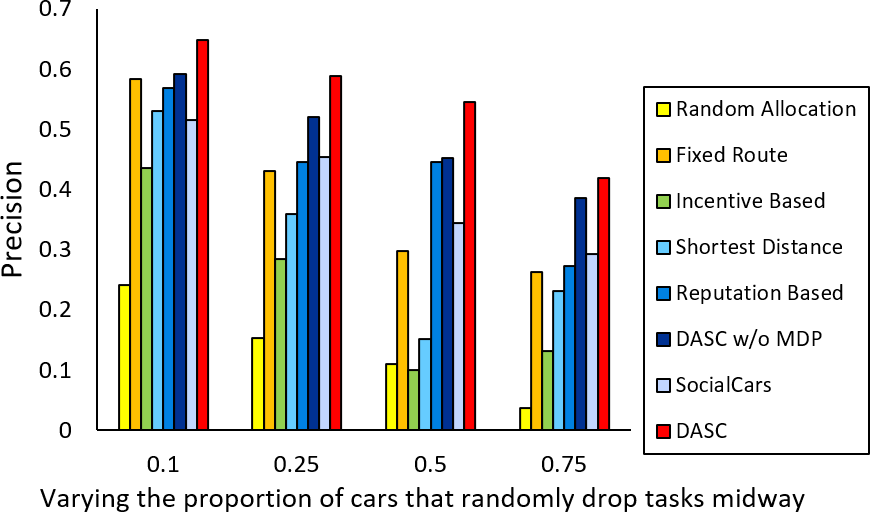}
    \vspace{-0.25in}
    \caption{Precision vs. Proportion of car drivers that randomly drop tasks}
    \label{fig:Precision2}
  \end{minipage}
  \hfill
  \begin{minipage}[b]{0.3\textwidth}
    \includegraphics[width=5.7cm]{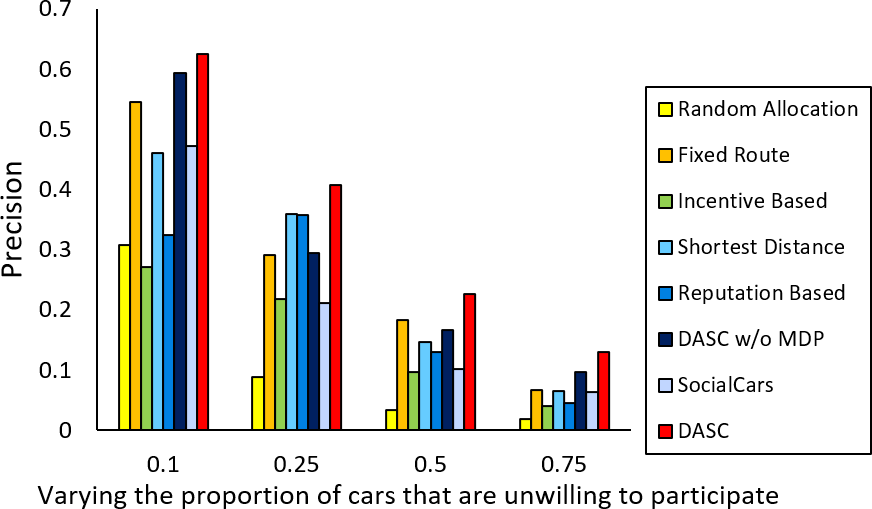}
    \vspace{-0.25in}
    \caption{Precision vs. Proportion of car drivers that are unwilling to participate}
    \label{fig:Precision3}
  \end{minipage}
\end{figure}

\begin{figure}[!h]
  \centering
  \begin{minipage}[b]{0.3\textwidth}
    \includegraphics[width=5.7cm]{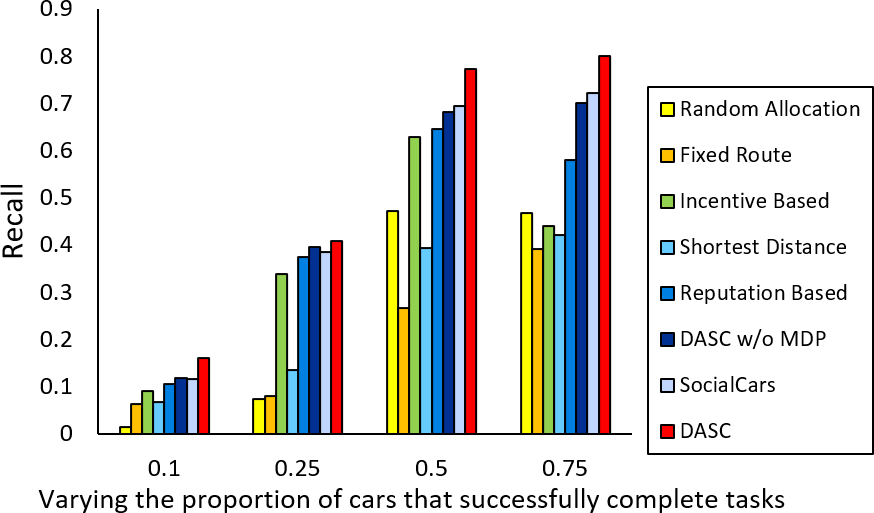}
    \vspace{-0.25in}
    \caption{Recall vs. Proportion of car drivers that successfully complete tasks}
    \label{fig:Recall1}
  \end{minipage}
  \hfill
  \begin{minipage}[b]{0.3\textwidth}
    \includegraphics[width=5.7cm]{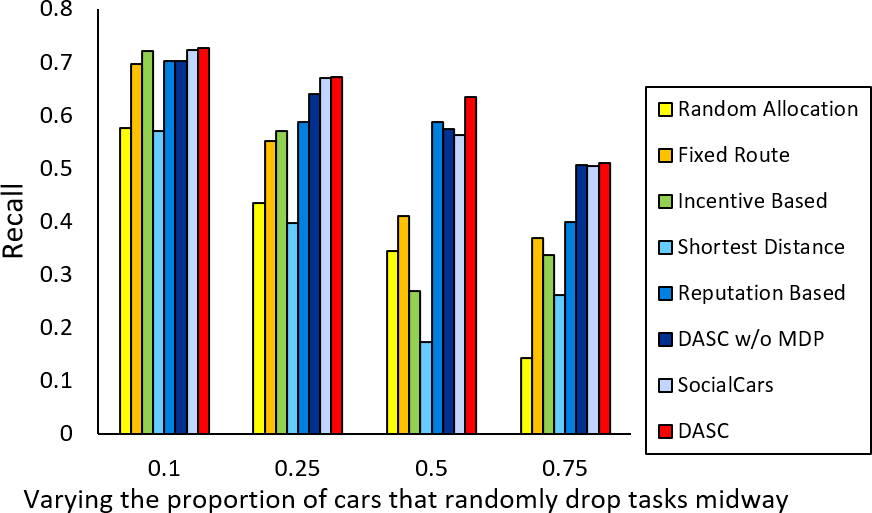}
    \vspace{-0.25in}
    \caption{Recall vs. Proportion of car drivers that randomly drop tasks}
    \label{fig:Recall2}
  \end{minipage}
  \hfill
  \begin{minipage}[b]{0.3\textwidth}
    \includegraphics[width=5.7cm]{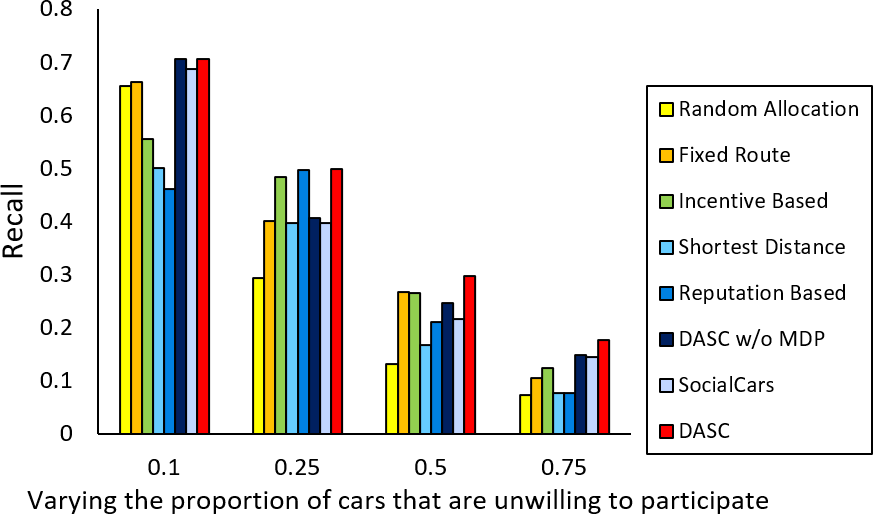}
    \vspace{-0.25in}
    \caption{Recall vs. Proportion of car drivers that are unwilling to participate}
    \label{fig:Recall3}
  \end{minipage}
\end{figure}

\begin{figure}[!h]
  \centering
  \begin{minipage}[b]{0.3\textwidth}
    \includegraphics[width=5.7cm]{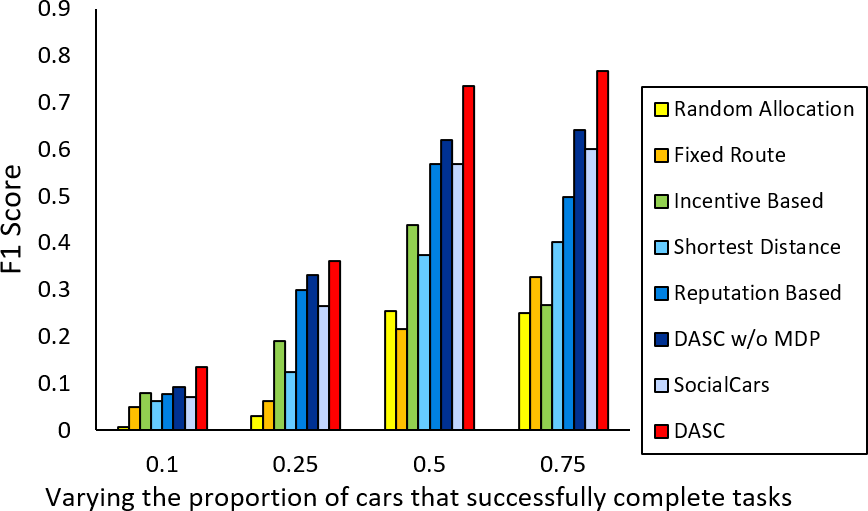}
    \vspace{-0.25in}
    \caption{F1 Score vs. Proportion of car drivers that successfully complete tasks}
    \label{fig:F1-1}
  \end{minipage}
  \hfill
  \begin{minipage}[b]{0.3\textwidth}
    \includegraphics[width=5.7cm]{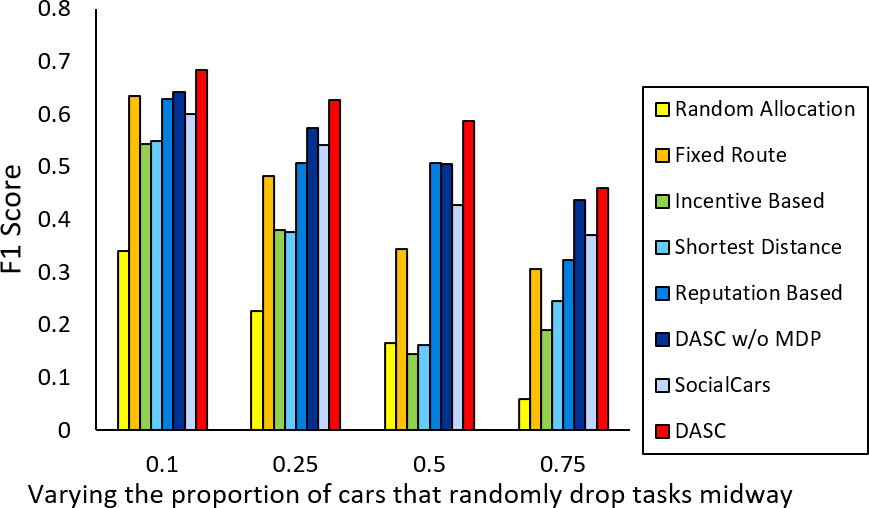}
    \vspace{-0.25in}
    \caption{F1 Score vs. Proportion of car drivers that randomly drop tasks}
    \label{fig:F1-2}
  \end{minipage}
  \hfill
  \begin{minipage}[b]{0.3\textwidth}
    \includegraphics[width=5.7cm]{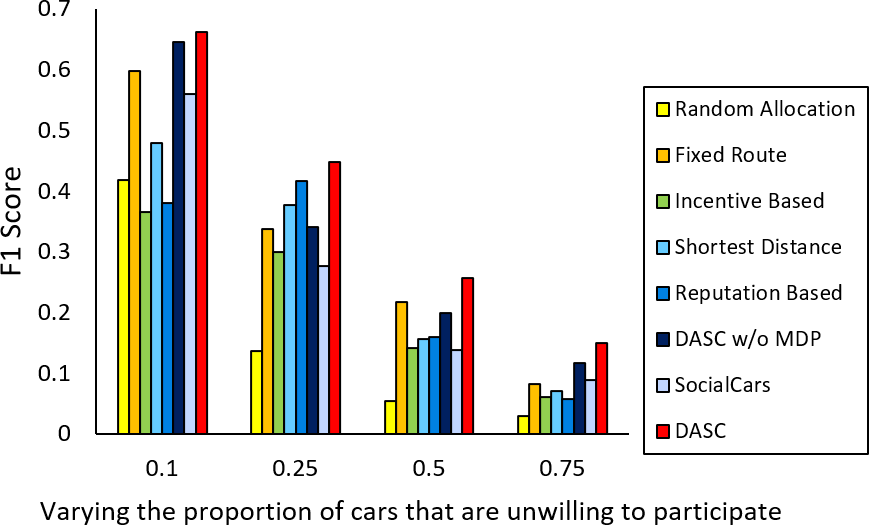}
    \vspace{-0.25in}
    \caption{F1 Score vs. Proportion of car drivers that are unwilling to participate}
    \label{fig:F1-3}
  \end{minipage}
\end{figure}

For each experiment, at any given time we vary the proportion of cars across one category and equally distribute the rest of the cars across the other two categories. Figures~\ref{fig:Deadlines2}-\ref{fig:Deadlines4} show the deadline hit rate while varying the proportions of car drivers that i) successfully complete tasks, ii) randomly drop tasks, and ii) are unwilling to participate, respectively. Likewise, Figures~\ref{fig:Accuracy1}-\ref{fig:Accuracy3} illustrate the corresponding accuracy, Figures~\ref{fig:Precision1}-\ref{fig:Precision3} illustrates the corresponding precision, Figures~\ref{fig:Recall1}-\ref{fig:Recall3} illustrate the corresponding recall, and Figures~\ref{fig:F1-1}-\ref{fig:F1-3} illustrates the corresponding F1 scores while varying the three types of drivers. We note that DASC achieves the highest deadline hit rate, accuracy, precision, recall, and F1 scores when the combination of cars across different categories changes. This improvement is primarily attributed to the dynamic incentive control (DIC) module that helps to adjust the rewards considering the aggregate reputation of the cars. If the aggregate reputation for a task falls, the module increases the reward which encourages the cars to select the task. This implicitly ensures the selection of the maximum number of tasks within their given deadlines. The performance gain is also imputed to the road damage discovery (RDD) module that sends out the scout cars for locating the road damage, which in turn helps to make better routing decisions by the vehicle dispatch (VD) module.

	\section{Discussion}\label{sec:discussion}
DASC is designed to only operate in \emph{post-disaster} scenarios but not during the disaster itself and in safe environments where at least some road networks are operational allowing cars to be able to traverse safely along them. The priority of DASC is to ensure the safety of the participating car drivers and thus DASC is not intended to operate in life-threatening environments where the lives of car drivers are at risk (e.g., during a nuclear explosion).

The DASC framework is also designed to be a general-purpose sensing response framework that can not only operate in the aftermath of disaster scenarios but could be seamlessly extended to other sensing applications. \mt{For example, the DASC can be applied to smart city applications such as urban noise mapping, detecting air pollution, free parking spot locating, traffic congestion detection.} Depending on the application scenario, the search criteria for the Tweets can be modified. \mt{For example, in an urban noise mapping application, the SSD module can be adapted to look for events indicating ``noisy streets" instead of ``pharmacy availability". Furthermore, the DIC module's sensitivity to the dropping of tasks can be adjusted by tuning the PID constants.} For example, in a free parking spot locating application, the rewards adjustment can be made less sensitive. Likewise, the Utility Function on the VD module can include additional factors for assigning the tasks to the cars. For example, the speed of the cars can be incorporated in the Utility Function in a traffic congestion detection application.
	
The flexibility of DASC allows it to be extended to incorporate information from multiple sources beyond cars. Based on the application scenario, the cars can be complemented with other information sources such as weather databases, official news agencies, field agents (i.e., rescuers, firefighters), crowdsourcing, etc. The DASC framework could be enhanced by incorporating the additional sensing information into the Vehicle Dispatch (VD) Module. Specifically, the Utility Function in the VD module can be modified accordingly to use the additional sensing data as a task allocation factor to lower dependencies on the car. For example, if there is an additional input signal from official news agencies, Equation \ref{eq:priority} can be updated with a fourth factor indicating the assertion of the news agency along with its weight (e.g., $\lambda_4$). The news agency can be given greater credibility than the event veracity score from social media posts. Moreover, depending on the scenario, the DASC framework could be able to conclude the actual truth of certain events without dispatching cars, thereby saving resources that can be utilized elsewhere to cover other events.

We acknowledge that it is likely that the network connectivity could be lost during a disaster. The basis for DASC's successful operation is to have network connectivity between the car drivers and the framework's backend so that they may establish communication. Therefore, DASC is intended to be only operated post-disaster scenarios and safe environments where some form of network connectivity is present. It is reasonable to assume that after a disaster at least some degree of cellular or Wi-Fi connectivity could be available depending on the type of the disaster~\cite{casoni2015integration}. The DASC framework can also be enhanced to handle the temporal loss of connectivity (i.e., intermittent) and sparse connectivity. For example, in applications where an intermittent network loss may occur, a layered vehicular delay-tolerant network (DTN) can be implemented through DTN gateways on proximal vehicles (i.e., nearby cars) to provide persistent storage for storing sensing data~\cite{soares2009layered}. Moreover, in applications involving spotty connectivity, a multi-hop wireless mesh network could be established between the vehicles to strengthen areas with weaker signal strength~\cite{draves2004routing}.
	\section{Conclusion}
In this paper, we develop the DASC scheme for a road damage-aware social-media-driven car sensing framework in reliable SCS applications. DASC addresses three intrinsic challenges in integrating the social media with cars: i) utilizing unreliable social signals to drive cars to reported event locations; ii) mitigating the adverse effect of the churn introduced by the rational car drivers; and iii) handling the road-damages caused by disasters to optimally guide cars to destinations. The results from a rigorous evaluation with a real-world disaster recovery case study reveal that the DASC achieves remarkable performance gains over the state-of-the-art VSNs-based sensing systems. We envision the outcomes of this paper to pave the road for a novel road damage-aware social-media-driven car sensing system to expedite the recovery phases of unexpected calamities.
	
\section*{Acknowledgment}	
This research is supported in part by the National Science Foundation under Grant No. CNS-1845639, CNS-1831669, Army Research Office under Grant W911NF-17-1-0409. The views and conclusions contained in this document are those of the authors and should not be interpreted as representing the official policies, either expressed or implied, of the Army Research Office or the U.S. Government. The U.S. Government is authorized to reproduce and distribute reprints for Government purposes notwithstanding any copyright notation here on.

\bibliographystyle{elsarticle-num}
\bibliography{ref.bib}

\begin{thebibliography}{10}
\expandafter\ifx\csname url\endcsname\relax
  \def\url#1{\texttt{#1}}\fi
\expandafter\ifx\csname urlprefix\endcsname\relax\def\urlprefix{URL }\fi
\expandafter\ifx\csname href\endcsname\relax
  \def\href#1#2{#2} \def\path#1{#1}\fi

\bibitem{zhang2008efficient}
C.~Zhang, R.~Lu, X.~Lin, P.-H. Ho, X.~Shen, An efficient identity-based batch
  verification scheme for vehicular sensor networks, in: IEEE INFOCOM 2008-The
  27th Conference on Computer Communications, IEEE, 2008, pp. 246--250 (2008).

\bibitem{park2016motives}
S.~Park, J.~Kim, R.~Mizouni, U.~Lee, Motives and concerns of dashcam video
  sharing, in: Proceedings of the 2016 CHI Conference on Human Factors in
  Computing Systems, ACM, 2016, pp. 4758--4769 (2016).

\bibitem{wang2019age}
D.~Wang, B.~K. Szymanski, T.~Abdelzaher, H.~Ji, L.~Kaplan, The age of social
  sensing, Computer 52~(1) (2019) 36--45 (2019).

\bibitem{zhang2018real}
D.~Zhang, Y.~Ma, Y.~Zhang, S.~Lin, X.~S. Hu, D.~Wang, A real-time and
  non-cooperative task allocation framework for social sensing applications in
  edge computing systems, in: 2018 IEEE Real-Time and Embedded Technology and
  Applications Symposium (RTAS), IEEE, 2018, pp. 316--326 (2018).

\bibitem{noulas2012tale}
A.~Noulas, S.~Scellato, R.~Lambiotte, M.~Pontil, C.~Mascolo, A tale of many
  cities: universal patterns in human urban mobility, PloS one 7~(5) (2012)
  e37027 (2012).

\bibitem{zhang2017constraint}
D.~Y. Zhang, D.~Wang, Y.~Zhang, Constraint-aware dynamic truth discovery in big
  data social media sensing, in: Big Data (Big Data), 2017 IEEE International
  Conference on, IEEE, 2017, pp. 57--66 (2017).

\bibitem{wang2015social}
D.~Wang, T.~Abdelzaher, L.~Kaplan, Social sensing: building reliable systems on
  unreliable data, Morgan Kaufmann, 2015 (2015).

\bibitem{lichtveld2018disasters}
M.~Lichtveld, Disasters through the lens of disparities: elevate community
  resilience as an essential public health service (2018).

\bibitem{zhang2019crowdlearn}
D.~Zhang, Y.~Zhang, Q.~Li, T.~Plummer, D.~Wang, Crowdlearn: A crowd-ai hybrid
  system for deep learning-based damage assessment applications, in: 2019 IEEE
  39th International Conference on Distributed Computing Systems (ICDCS), IEEE,
  2019, pp. 1221--1232 (2019).

\bibitem{wang2019social}
D.~Wang, D.~Zhang, Y.~Zhang, M.~T. Rashid, L.~Shang, N.~Wei, Social edge
  intelligence: Integrating human and artificial intelligence at the edge, in:
  2019 IEEE First International Conference on Cognitive Machine Intelligence
  (CogMI), IEEE, 2019, pp. 194--201 (2019).

\bibitem{zhang2016robust}
D.~Y. Zhang, R.~Han, D.~Wang, C.~Huang, On robust truth discovery in sparse
  social media sensing, in: 2016 IEEE International Conference on Big Data (Big
  Data), IEEE, 2016, pp. 1076--1081 (2016).

\bibitem{rhea2004handling}
S.~Rhea, D.~Geels, T.~Roscoe, J.~Kubiatowicz, et~al., Handling churn in a dht,
  in: Proceedings of the USENIX Annual Technical Conference, Vol.~6, Boston,
  MA, USA, 2004, pp. 127--140 (2004).

\bibitem{vance2019towards}
N.~Vance, M.~T. Rashid, D.~Zhang, D.~Wang, Towards reliability in online
  high-churn edge computing: A deviceless pipelining approach, in: 2019 IEEE
  International Conference on Smart Computing (SMARTCOMP), IEEE, 2019, pp.
  301--308 (2019).

\bibitem{zhang2019heteroedge}
D.~Zhang, T.~Rashid, X.~Li, N.~Vance, D.~Wang, Heteroedge: taming the
  heterogeneity of edge computing system in social sensing, in: Proceedings of
  the International Conference on Internet of Things Design and Implementation,
  2019, pp. 37--48 (2019).

\bibitem{godfrey2006minimizing}
P.~Godfrey, S.~Shenker, I.~Stoica, Minimizing churn in distributed systems,
  Vol.~36, ACM, 2006 (2006).

\bibitem{haeberlen2006fallacies}
A.~Haeberlen, A.~Mislove, A.~Post, P.~Druschel, Fallacies in evaluating
  decentralized systems., in: IPTPS, 2006, pp. 1--6 (2006).

\bibitem{zhang2010cloud}
Q.~Zhang, L.~Cheng, R.~Boutaba, Cloud computing: state-of-the-art and research
  challenges, Journal of internet services and applications 1~(1) (2010) 7--18
  (2010).

\bibitem{bono2011network}
F.~Bono, E.~Guti{\'e}rrez, A network-based analysis of the impact of structural
  damage on urban accessibility following a disaster: the case of the
  seismically damaged port au prince and carrefour urban road networks, Journal
  of Transport Geography 19~(6) (2011) 1443--1455 (2011).

\bibitem{baker2003genetic}
B.~M. Baker, M.~Ayechew, A genetic algorithm for the vehicle routing problem,
  Computers \& Operations Research 30~(5) (2003) 787--800 (2003).

\bibitem{prins2004simple}
C.~Prins, A simple and effective evolutionary algorithm for the vehicle routing
  problem, Computers \& Operations Research 31~(12) (2004) 1985--2002 (2004).

\bibitem{rashid2019socialcar}
M.~T. Rashid, D.~Zhang, D.~Wang, Socialcar: A task allocation framework for
  social media driven vehicular network sensing systems, in: The 15th
  International Conference on Mobile Ad-hoc and Sensor Networks (MSN), IEEE,
  2019, pp. 125--130 (2019).

\bibitem{nekovee2005sensor}
M.~Nekovee, Sensor networks on the road: the promises and challenges of
  vehicular ad hoc networks and grids, in: Workshop on ubiquitous computing and
  e-Research, Vol.~47, 2005, pp. 1--6 (2005).

\bibitem{lee2006mobeyes}
U.~Lee, B.~Zhou, M.~Gerla, E.~Magistretti, P.~Bellavista, A.~Corradi, Mobeyes:
  smart mobs for urban monitoring with a vehicular sensor network, IEEE
  Wireless Communications 13~(5) (2006).

\bibitem{galeso2016waze}
M.~Galeso, Waze: An Easy Guide to the Best Features, Lulu Press, Inc, 2016
  (2016).

\bibitem{dong2008automatic}
Y.~F. Dong, S.~Kanhere, C.~T. Chou, N.~Bulusu, Automatic collection of fuel
  prices from a network of mobile cameras, in: DCoSS 2008, Springer, 2008, pp.
  140--156 (2008).

\bibitem{IPSN:12}
D.~Wang, L.~Kaplan, H.~Le, T.~Abdelzaher, On truth discovery in social sensing:
  A maximum likelihood estimation approach, in: Proc. ACM/IEEE 11th Int
  Information Processing in Sensor Networks (IPSN) Conf, 2012, pp. 233--244
  (Apr. 2012).
\newblock \href {https://doi.org/10.1109/IPSN.2012.6920960}
  {\path{doi:10.1109/IPSN.2012.6920960}}.

\bibitem{wang2014using}
D.~Wang, M.~T. Amin, S.~Li, T.~Abdelzaher, L.~Kaplan, S.~Gu, C.~Pan, H.~Liu,
  C.~C. Aggarwal, R.~Ganti, Using humans as sensors: an estimation-theoretic
  perspective, in: Information Processing in Sensor Networks, IPSN-14
  Proceedings of the 13th International Symposium on, IEEE, 2014, pp. 35--46
  (2014).

\bibitem{zhang2020graphcast}
Y.~Zhang, X.~Dong, L.~Shang, D.~Zhang, D.~Wang, A multi-modal graph neural
  network approach to traffic risk forecasting in smart urban sensing, in:
  International Conference on Sensing, Communication, and Networking (SECON),
  IEEE, 2020, p. to appear (2020).

\bibitem{al2014crowd}
M.~T. Al~Amin, T.~Abdelzaher, D.~Wang, B.~Szymanski, Crowd-sensing with
  polarized sources, in: 2014 IEEE International Conference on Distributed
  Computing in Sensor Systems, IEEE, 2014, pp. 67--74 (2014).

\bibitem{marshall2016mood}
J.~Marshall, D.~Wang, Mood-sensitive truth discovery for reliable
  recommendation systems in social sensing, in: Proceedings of International
  Conference on Recommender Systems (Recsys), ACM, 2016 (2016).

\bibitem{wang2013recursive}
D.~Wang, T.~Abdelzaher, L.~Kaplan, C.~C. Aggarwal, Recursive fact-finding: A
  streaming approach to truth estimation in crowdsourcing applications, in:
  2013 IEEE 33rd International Conference on Distributed Computing Systems,
  IEEE, 2013, pp. 530--539 (2013).

\bibitem{zhang2019sparse}
D.~Zhang, Y.~Zhang, Q.~Li, D.~Wang, Sparse user check-in venue prediction by
  exploring latent decision contexts from location-based social networks, IEEE
  Transactions on Big Data (2019).

\bibitem{zhang2017large}
D.~Y. Zhang, D.~Wang, H.~Zheng, X.~Mu, Q.~Li, Y.~Zhang, Large-scale
  point-of-interest category prediction using natural language processing
  models, in: 2017 IEEE International Conference on Big Data (Big Data), IEEE,
  2017, pp. 1027--1032 (2017).

\bibitem{zhang2019edgebatch}
D.~Zhang, N.~Vance, Y.~Zhang, M.~T. Rashid, D.~Wang, Edgebatch: Towards
  ai-empowered optimal task batching in intelligent edge systems, in: 2019 IEEE
  Real-Time Systems Symposium (RTSS), IEEE, 2019, pp. 366--379 (2019).

\bibitem{wang2013credibility}
D.~Wang, L.~Kaplan, T.~Abdelzaher, C.~C. Aggarwal, On credibility estimation
  tradeoffs in assured social sensing, IEEE Journal on Selected Areas in
  Communications 31~(6) (2013) 1026--1037 (2013).

\bibitem{wang2012scalability}
D.~Wang, L.~Kaplan, T.~Abdelzaher, C.~C. Aggarwal, On scalability and
  robustness limitations of real and asymptotic confidence bounds in social
  sensing, in: 2012 9th Annual IEEE Communications Society Conference on
  Sensor, Mesh and Ad Hoc Communications and Networks (SECON), IEEE, 2012, pp.
  506--514 (2012).

\bibitem{zhang2019deeprisk}
Y.~Zhang, H.~Wang, D.~Zhang, D.~Wang, Deeprisk: A deep transfer learning
  approach to migratable traffic risk estimation in intelligent transportation
  using social sensing, in: 2019 15th International Conference on Distributed
  Computing in Sensor Systems (DCOSS), IEEE, 2019, pp. 123--130 (2019).

\bibitem{vance2018privacy}
N.~Vance, D.~Y. Zhang, Y.~Zhang, D.~Wang, Privacy-aware edge computing in
  social sensing applications using ring signatures, in: 2018 IEEE 24th
  International Conference on Parallel and Distributed Systems (ICPADS), IEEE,
  2018, pp. 755--762 (2018).

\bibitem{zhang2019integrated}
D.~Y. Zhang, D.~Wang, An integrated top-down and bottom-up task allocation
  approach in social sensing based edge computing systems, in: IEEE INFOCOM
  2019-IEEE Conference on Computer Communications, IEEE, 2019, pp. 766--774
  (2019).

\bibitem{xu2016participatory}
Z.~Xu, H.~Zhang, V.~Sugumaran, K.-K.~R. Choo, L.~Mei, Y.~Zhu, Participatory
  sensing-based semantic and spatial analysis of urban emergency events using
  mobile social media, EURASIP Journal on Wireless Communications and
  Networking 2016~(1) (2016) 44 (2016).

\bibitem{chen2014road}
P.-T. Chen, F.~Chen, Z.~Qian, Road traffic congestion monitoring in social
  media with hinge-loss markov random fields, in: 2014 IEEE International
  Conference on Data Mining, IEEE, 2014, pp. 80--89 (2014).

\bibitem{imran2014aidr}
M.~Imran, C.~Castillo, J.~Lucas, P.~Meier, S.~Vieweg, Aidr: Artificial
  intelligence for disaster response, in: Proceedings of the 23rd International
  Conference on World Wide Web, ACM, 2014, pp. 159--162 (2014).

\bibitem{wang2014provenance}
D.~Wang, M.~T. Al~Amin, T.~Abdelzaher, D.~Roth, C.~R. Voss, L.~M. Kaplan,
  S.~Tratz, J.~Laoudi, D.~Briesch, Provenance-assisted classification in social
  networks, IEEE Journal of Selected Topics in Signal Processing 8~(4) (2014)
  624--637 (2014).

\bibitem{wang2014surrogate}
D.~Wang, T.~Abdelzaher, L.~Kaplan, Surrogate mobile sensing, IEEE
  Communications Magazine 52~(8) (2014) 36--41 (2014).

\bibitem{zhang2019social}
D.~Zhang, N.~Vance, D.~Wang, When social sensing meets edge computing: Vision
  and challenges, in: 2019 28th International Conference on Computer
  Communication and Networks (ICCCN), IEEE, 2019, pp. 1--9 (2019).

\bibitem{rashid2019collabdrone}
M.~T. Rashid, D.~Zhang, Z.~Liu, H.~Lin, D.~Wang, Collabdrone: A collaborative
  spatiotemporal-aware drone sensing system driven by social sensing signals,
  in: 2019 28th International Conference on Computer Communication and Networks
  (ICCCN), IEEE, 2019, pp. 1--9 (2019).

\bibitem{rashid2019sead}
M.~T. Rashid, D.~Y. Zhang, L.~Shang, D.~Wang, Sead: Towards a
  social-media-driven energy-aware drone sensing framework, in: 2019 IEEE 25th
  International Conference on Parallel and Distributed Systems (ICPADS), IEEE,
  2019, pp. 647--654 (2019).

\bibitem{gao2015survey}
H.~Gao, C.~H. Liu, W.~Wang, J.~Zhao, Z.~Song, X.~Su, J.~Crowcroft, K.~K. Leung,
  A survey of incentive mechanisms for participatory sensing, IEEE
  Communications Surveys \& Tutorials 17~(2) (2015) 918--943 (2015).

\bibitem{hsueh2008dynamic}
C.-F. Hsueh, H.-K. Chen, H.-W. Chou, Dynamic vehicle routing for relief
  logistics in natural disasters, in: Vehicle routing problem, IntechOpen,
  2008, pp. 71--84 (2008).

\bibitem{korkmaz2016path}
S.~A. Korkmaz, M.~Poyraz, Path planning for rescue vehicles via segmented
  satellite disaster images and gps road map, in: 2016 CISP-BMEI, IEEE, 2016,
  pp. 145--150 (2016).

\bibitem{mahmoudabadi2014solving}
A.~Mahmoudabadi, S.~M. Seyedhosseini, Solving hazmat routing problem in chaotic
  damage severity network under emergency environment, Transport policy 36
  (2014) 34--45 (2014).

\bibitem{kuntze2012seneka}
H.-B. Kuntze, C.~W. Frey, I.~Tchouchenkov, B.~Staehle, E.~Rome, K.~Pfeiffer,
  A.~Wenzel, J.~W{\"o}llenstein, Seneka-sensor network with mobile robots for
  disaster management, in: 2012 IEEE Conference on Technologies for Homeland
  Security (HST), IEEE, 2012, pp. 406--410 (2012).

\bibitem{singh2018analyzing}
N.~Singh, N.~Roy, A.~Gangopadhyay, Analyzing the sentiment of crowd for
  improving the emergency response services, in: 2018 IEEE International
  Conference on Smart Computing (SMARTCOMP), IEEE, 2018, pp. 1--8 (2018).

\bibitem{zhang2020pqa}
Y.~Zhang, X.~Dong, M.~T. Rashid, L.~Shang, J.~Han, D.~Zhang, D.~Wang, Pqa-cnn:
  Towards perceptual quality assured single-image super-resolution in remote
  sensing, in: The 17th Annual IEEE Communications Society Conference on
  Sensor, Mesh and Ad Hoc Communications and Networks (SECON 2020), IEEE, 2020
  (2020).

\bibitem{mokhtari2019sliding}
F.~Mokhtari, M.~I. Akhlaghi, S.~L. Simpson, G.~Wu, P.~J. Laurienti, Sliding
  window correlation analysis: Modulating window shape for dynamic brain
  connectivity in resting state, NeuroImage 189 (2019) 655--666 (2019).

\bibitem{duchovn2014path}
F.~Ducho{\v{n}}, A.~Babinec, M.~Kajan, P.~Be{\v{n}}o, M.~Florek, T.~Fico,
  L.~Juri{\v{s}}ica, Path planning with modified a star algorithm for a mobile
  robot, Procedia Engineering 96 (2014) 59--69 (2014).

\bibitem{rashid2020socialdrone}
M.~T. Rashid, D.~Zhang, L.~Shang, D.~Wang, An integrated social media and drone
  sensing system for reliable disaster response, in: IEEE INFOCOM 2020-IEEE
  Conference on Computer Communications, IEEE, 2020 (2020).

\bibitem{ieong2005fast}
S.~Ieong, R.~McGrew, E.~Nudelman, Y.~Shoham, Q.~Sun, Fast and compact: A simple
  class of congestion games, in: AAAI, Vol.~5, 2005, pp. 489--494 (2005).

\bibitem{doyle2013feedback}
J.~C. Doyle, B.~A. Francis, A.~R. Tannenbaum, Feedback control theory, Courier
  Corporation, 2013 (2013).

\bibitem{rashid2020compdrone}
M.~T. Rashid, Y.~Zhang, D.~Y. Zhang, D.~Wang, Compdrone: Towards integrated
  computational model and social drone based wildfire monitoring, in: 16th
  International Conference on Distributed Computing in Sensor Systems,
  (DCOSS20), IEEE, 2020, accepted (2020).

\bibitem{even2009online}
E.~Even-Dar, S.~M. Kakade, Y.~Mansour, Online markov decision processes,
  Mathematics of Operations Research 34~(3) (2009) 726--736 (2009).

\bibitem{geisberger2008contraction}
R.~Geisberger, P.~Sanders, D.~Schultes, D.~Delling, Contraction hierarchies:
  Faster and simpler hierarchical routing in road networks, in: International
  Workshop on Experimental and Efficient Algorithms, Springer, 2008, pp.
  319--333 (2008).

\bibitem{raykar2014sequential}
V.~Raykar, P.~Agrawal, Sequential crowdsourced labeling as an epsilon-greedy
  exploration in a markov decision process, in: Artificial intelligence and
  statistics, 2014, pp. 832--840 (2014).

\bibitem{dosovitskiy2017carla}
A.~Dosovitskiy, G.~Ros, F.~Codevilla, A.~Lopez, V.~Koltun, Carla: An open urban
  driving simulator, arXiv preprint arXiv:1711.03938 (2017).

\bibitem{haklay2008openstreetmap}
M.~Haklay, P.~Weber, Openstreetmap: User-generated street maps, IEEE Pervasive
  Computing 7~(4) (2008) 12--18 (2008).

\bibitem{joshi2016accuracy}
R.~Joshi, Accuracy, precision, recall \& f1 score: Interpretation of
  performance measures, Retrieved April 1 (2016) 2018 (2016).

\bibitem{floudas1995nonlinear}
C.~A. Floudas, Nonlinear and mixed-integer optimization: fundamentals and
  applications, Oxford University Press, 1995 (1995).

\bibitem{luersen2004globalized}
M.~A. Luersen, R.~Le~Riche, Globalized nelder--mead method for engineering
  optimization, Computers \& structures 82~(23-26) (2004) 2251--2260 (2004).

\bibitem{pilanci2017newton}
M.~Pilanci, M.~J. Wainwright, Newton sketch: A near linear-time optimization
  algorithm with linear-quadratic convergence, SIAM Journal on Optimization
  27~(1) (2017) 205--245 (2017).

\bibitem{murphy2017harvey}
R.~Murphy, Where harvey’s effects were felt the most in texas, Online:
  https://apps.texastribune.org/harvey-fema-damage-analysis/ (2017).

\bibitem{moran2015speeding}
\href{http://www.fleetopenclosed.org/}{Multi-state fleet response working group
  (fwg) report} (2017).
\newline\urlprefix\url{http://www.fleetopenclosed.org/}

\bibitem{portugal2010msp}
D.~Portugal, R.~Rocha, Msp algorithm: multi-robot patrolling based on territory
  allocation using balanced graph partitioning, in: Proceedings of the 2010 ACM
  symposium on applied computing, ACM, 2010, pp. 1271--1276 (2010).

\bibitem{palazzi2012delay}
C.~E. Palazzi, F.~Pezzoni, P.~M. Ruiz, Delay-bounded data gathering in urban
  vehicular sensor networks, Pervasive and Mobile Computing 8~(2) (2012)
  180--193 (2012).

\bibitem{gong2014social}
H.~Gong, L.~Yu, X.~Zhang, Social contribution-based routing protocol for
  vehicular network with selfish nodes, International Journal of Distributed
  Sensor Networks 10~(4) (2014) 753024 (2014).

\bibitem{pasadena}
\href{https://kristinamorales.com/guides/pasadena-tx/}{Pasadena texas city
  guide} (2020).
\newline\urlprefix\url{https://kristinamorales.com/guides/pasadena-tx/}

\bibitem{wilde1976social}
G.~J. Wilde, Social interaction patterns in driver behavior: An introductory
  review, Human factors 18~(5) (1976) 477--492 (1976).

\bibitem{wang2014modeling}
W.~Wang, J.~Xi, H.~Chen, Modeling and recognizing driver behavior based on
  driving data: A survey, Mathematical Problems in Engineering 2014 (2014).

\bibitem{casoni2015integration}
M.~Casoni, C.~A. Grazia, M.~Klapez, N.~Patriciello, A.~Amditis, E.~Sdongos,
  Integration of satellite and lte for disaster recovery, IEEE Communications
  Magazine 53~(3) (2015) 47--53 (2015).

\bibitem{soares2009layered}
V.~N. Soares, F.~Farahmand, J.~J. Rodrigues, A layered architecture for
  vehicular delay-tolerant networks, in: 2009 IEEE Symposium on Computers and
  Communications, IEEE, 2009, pp. 122--127 (2009).

\bibitem{draves2004routing}
R.~Draves, J.~Padhye, B.~Zill, Routing in multi-radio, multi-hop wireless mesh
  networks, in: Proceedings of the 10th annual international conference on
  Mobile computing and networking, 2004, pp. 114--128 (2004).

\end{thebibliography}

\end{document}